\documentclass{revtex4}

\usepackage{amsthm,amssymb,amsmath}				
\usepackage{graphicx,comment}
\usepackage{float}   
\usepackage{xcolor}

\begin{document}

\title{Massless KG-oscillators in Som-Raychaudhuri cosmic string spacetime in a fine tuned rainbow gravity }
\author{Omar Mustafa}
\email{omar.mustafa@emu.edu.tr}
\affiliation{Department of Physics, Eastern Mediterranean University, G. Magusa, north
Cyprus, Mersin 10 - Turkiye.}

\begin{abstract}
\textbf{Abstract:}\
A fine tuned rainbow gravity describes both relativistic quantum particles and anti-particles alike. That is, the ratio $y=E/E_{P}$ in the rainbow functions $g_{_{0}}\left( y\right) $ and $%
g_{_{1}}\left( y\right) $ should be fine tuned into $0\leq y=E/E_{P}\leq
1\Rightarrow y=\left\vert E\right\vert /E_{P}$, otherwise rainbow gravity will only secure Planck's energy scale $E_p$ invariance for relativistic particles and the anti-particles are left unfortunate (in the sense that their energies will be indefinitely unbounded). Using this fine tuning we discuss the rainbow gravity effect on Klein-Gordon (KG) oscillators in Som-Raychaudhuri cosmic string rainbow gravity spacetime background. We use the rainbow functions: (i) $g_{_{0}}\left( y\right) =1$, $%
g_{_{1}}\left( y\right) =\sqrt{1-\epsilon y^{n}}, n=0,1$, loop quantum gravity motivated pairs,  (ii)  $%
g_{_{0}}\left( y\right) =g_{_{1}}\left( y\right) =\left( 1-\epsilon y\right)
^{-1}$, a horizon problem motivated pair, and (iii) $g_{_{0}}\left( y\right) =\left( e^{\epsilon y}-1\right)
/\epsilon y$, $g_{_{1}}\left( y\right) =1$,  a gamma-ray bursts motivated pair. We show that the energies obtained using the first two rainbow functions in (i) completely comply with the rainbow gravity model (on the invariance of the Planck's energy scale $E_{p}$). The rainbow function pair in (ii) has no effect on massless KG-oscillators. Whereas, the one in (iii) does not show any eminent tendency towards the invariance of the Planck's energy scale. Yet, we suggest a new rainbow function pair $(g_0(y)=(1-\epsilon y)^{-1}, g_1 (y)=1)$,  and show that it secures invariance of the Planck's energy scale $E_p$. Moreover, similar performance is observed when this new pair is used for KG-oscillators and KG-Coulombic particles in cosmic string rainbow gravity spacetime and magnetic fields.


\end{abstract}

\maketitle

\section{Introduction}

A common suggestion of most approaches to quantum gravity (e.g., string field theory \cite{Re1}, loop quantum gravity \cite{Re2}, and non-commutative geometry \cite{Re3}) is to modify the standard relativistic energy-momentum dispersion relation in the ultraviolet regime into%
\begin{equation}
E^{2}g_{_{0}}\left( y\right) ^{2}-p^{2}c^{2}g_{_{1}}\left( y\right)
^{2}=m^{2}c^{4};\;y=E/E_{p},  \label{eq0.1}
\end{equation}%
where $g_{_{0}}\left( y\right) $, $g_{_{1}}\left( y\right) $ are the rainbow functions, $mc^{2}$ is its rest mass energy, and the condition that the
limit $y\rightarrow 0\Rightarrow g_{_{k}}\left( y\right) =1;\;k=0,1$, retrieves the standard energy-momentum dispersion relation in the infrared regime (i.e, the usual general relativity is recovered). The Einstein's field equations are consequently affected by such modification to read $G_{\mu\nu}(E/E_p)=8\pi G(E/E_p) T_{\mu\nu}(E/E_p) $, where $G(E/E_p)$ is, in this case, an energy-dependent Newton's universal gravitational constant that collapses into the conventional one $G=G(0)$ at the limit $E/E_p\rightarrow 0$. It should be interesting to mention that the modified dispersion relations are naturally produced by the doubly/deformed special relativity (DSR) \cite{Re4,Re5,Re6,Re7,Re8,Re9}. DSR extends special relativity and introduces yet another invariant energy scale (the Planck's energy $E_{p}$ as the maximum energy scale) alongside with the invariance of the speed of light. A generalization of DSR that includes curvature is provided by the doubly general relativity \cite%
{Re10}, where the spacetime metric depends on the energy of the probe particle and forms the so called rainbow gravity (RG).

Rainbow gravity has been an inviting research field over the years \cite%
{Re9,Re10,Re11,Re12,Re13,R7,R8,R81,R9,R10,R11,R12,R13,R14,R15,R16,R17}. For
example, the effects of RG are reported in the thermodynamical aspects of 
black holes \cite{Re12,Re13,R18,R19,R20,R21,R211,R221}, in the tests of
thresholds for ultra high-energy cosmic rays \cite{R13,R14,R15}, TeV photons 
\cite{R16}, gamma-ray bursts \cite{R7}, nuclear physics experiments \cite%
{R17}, dynamical stability conditions of neutron stars \cite{R22}, charged black holes in massive RG \cite{R222}, on geometrical thermodynamics and heat
engine of black holes in RG \cite{R223}, f(R) theories \cite{R224}, the initial singularity problem for closed rainbow cosmology \cite{R23}, the black hole entropy \cite{R24}, the removal of the singularity of the
early universe \cite{R25}, the Casimir effect in the rainbow Einstein's universe \cite{R8}, massive scalar field in RG Schwarzschild metric \cite%
{R26}, Yang--Mills black holes \cite{R27}, etc.

In different spacetime rainbow gravity backgrounds, on the other hand, recent studies are carried out on Klein-Gordon (KG) particles (i.e., spin-0 mesons), Dirac particles (spin-1/2 fermionic
particles), and Duffen-Kemmer-Peatiau (DKP) particles (spin-1 particles like bosons and photons). For example, in a cosmic string spacetime background in rainbow gravity, Landau levels via Schr\"{o}dinger and KG equations are studied by Bezzerra et al. \cite{R81}, Dirac oscillator and Aharonov-Bohm effect are studied by Bakke and Mota \cite{R28,R29}, DKP-particles by Hosseinpour et al. \cite{Re9}, quantum dynamics of photon by Sogut et al. \cite{R11}, KG-particles in a topologically trivial G\"{o}del-type spacetime by Kangal et al. \cite{R12}, etc. Recently, moreover, position-dependent mass (PDM) KG-particles in different spacetimes backgrounds are introduced \cite{R291,R292,R30,R31,R32,R33,R34,R35,R36}. It has been used in the study of PDM KG-Coulomb particles \cite{R291} in cosmic string rainbow gravity spacetime, PDM KG-oscillators in cosmic string spacetime within Kaluza-Klein theory \cite{R38}, in (2+1)-dimensional G\"{u}rses spacetime backgrounds \cite{R39}, and in Minkowski spacetime with space-like dislocation \cite{R40}. It could be interesting to mention that PDM is a notion used to describe coordinate transformation/deformation that renders the mass in the Schr\"{o}dinger equation to become effectively position-dependent
\cite{R31,R32,R33,R34,R35,R36}.

However, in our very recent study on PDM KG-Coulomb particles \cite{R291} in cosmic string rainbow gravity spacetime, we have followed the usual practice in the literature (e.g., \cite%
{Re9,Re10,Re12,R8,R81,R11,R12}) that $y=E/E_{p}$ in the rainbow functions (\ref{eq0.1}). This has led to some unfortunate results (except for one of the rainbow function models) for the KG-anti-particles (hence the notion that anti-particles are left unfortunate) as they did not satisfy the rainbow gravity model (i.e., their energies are reported to be $|E|>> E_{p}$). In the current methodical proposal, we use a necessary and convenient fine tuning on $y=E/E_{p}$ of the rainbow functions (\ref{eq0.1}) and constraint $%
y$ so that $0\leq y=E/E_{p}\leq 1$ for relativistic particles and anti-particles. This suggestion would allow us to write $y=\left\vert
E\right\vert /E_{p}$. We shall use this fine tuning and study massless KG-oscillators in a G\"{o}del-type Som-Raychaudhuri cosmic string rainbow gravity spacetime. We consider three pairs of rainbow functions: (a) $%
g_{_{0}}\left( y\right) =1$, $g_{_{1}}\left( y\right) =\sqrt{1-\epsilon y^{2}%
}$, and $g_{_{0}}\left( y\right) =1$, $g_{_{1}}\left( y\right) =\sqrt{%
1-\epsilon y}$, which are loop quantum gravity \cite{R41,R42} motivated pairs, (b) $g_{_{0}}\left( y\right) =g_{_{1}}\left(
y\right) =\left( 1-\epsilon y\right) ^{-1}$, a horizon problem \cite{Re5,R13} motivated pair, and (c) $g_{_{0}}\left( y\right) =\left(
e^{\epsilon y}-1\right) /\epsilon y$, $g_{_{1}}\left( y\right) =1$, motivated by gamma-ray bursts at cosmological distances \cite{Re6}. We propose yet a new experimental (metaphorically speaking) rainbow functions pair $g_{_{0}}\left( y\right) =\left( 1-\epsilon y\right) ^{-1},\, g_{_{1}}\left(
y\right) =1$ and study it for different KG-particles and anti-particles in different spacetime backgrounds. 

Hence, the Som-Raychaudhuri cosmic string spacetime metric (in the natural
units $c=\hbar =G=1$) 
\begin{equation}
ds^{2}=-\left( dt+\alpha \Omega r^{2}d\varphi \right) ^{2}+dr^{2}+\alpha
^{2}\,r^{2}d\varphi ^{2}+dz^{2},  \label{eq1}
\end{equation}%
under RG, takes the energy-dependent form%
\begin{equation}
ds^{2}=-\left( \frac{dt}{g_{_{0}}\left( y\right) }+\frac{\alpha \Omega
r^{2}d\varphi }{g_{_{1}}\left( y\right) }\right) ^{2}+\frac{1}{%
g_{_{1}}\left( y\right) ^{2}}\left[ dr^{2}+\alpha ^{2}\,r^{2}d\varphi
^{2}+dz^{2}\right] ;\;y=\left\vert E\right\vert /E_{p},  \label{eq2}
\end{equation}%
where $\alpha =1-4\mu $ is the deficit angle, $\mu $ is the linear mass density of the cosmic string so that $\alpha <1,E=E_{\pm }=\pm \left\vert
E\right\vert $ are the energies of the probe particle, $E_{+}=+\left\vert
E\right\vert $, and anti-particle, $E_{-}=-\left\vert E\right\vert $, in Som-Raychaudhuri cosmic string rainbow gravity spacetime, and $E_{p}=\sqrt{%
\hbar c^{5}/G}$ is the Planck energy. Here,the signature of the line elements (\ref{eq1}) and (\ref{eq2}) is $\left( -,+,+,+\right) $. Moreover,
the corresponding metric tensor $g_{\mu \nu }$ reads%
\begin{equation}
g_{\mu \nu }=\left( 
\begin{tabular}{cccc}
$\frac{-1}{g_{_{0}}\left( y\right) ^{2}}$ & $0$ & $\frac{-\alpha \Omega r^{2}%
}{g_{_{0}}\left( y\right) g_{_{1}}\left( y\right) }$ & $0\medskip $ \\ 
$0$ & $\frac{1}{g_{_{1}}\left( y\right) ^{2}}$ & $0$ & $0\medskip $ \\ 
$\frac{-\alpha \Omega r^{2}}{g_{_{0}}\left( y\right) g_{_{1}}\left( y\right) 
}$ & $0$ & $\frac{\left( \alpha ^{2}\,r^{2}-\alpha ^{2}\,\Omega
^{2}r^{4}\right) }{g_{_{1}}\left( y\right) ^{2}}$ & $0\medskip $ \\ 
$0$ & $0$ & $0$ & $\frac{1}{g_{_{1}}\left( y\right) ^{2}}\medskip $%
\end{tabular}%
\right) ;\;\mu ,\nu =t,r,\varphi ,z,  \label{eq3}
\end{equation}%
to imply 
\begin{equation*}
\det \left( g_{\mu \nu }\right) =g=-\frac{\alpha ^{2}\,r^{2}}{g_{_{0}}\left(
y\right) ^{2}g_{_{1}}\left( y\right) ^{6}},
\end{equation*}
and 
\begin{equation}
g^{\mu \nu }=\left( 
\begin{tabular}{cccc}
$-g_{_{0}}\left( y\right) ^{2}\left( 1-\Omega ^{2}r^{2}\right) $ & $0$ & $-%
\frac{\Omega g_{_{0}}\left( y\right) g_{_{1}}\left( y\right) }{\alpha }$ & $%
0\medskip $ \\ 
$0$ & $g_{_{1}}\left( y\right) ^{2}$ & $0$ & $0\medskip $ \\ 
$-\frac{\Omega g_{_{0}}\left( y\right) g_{_{1}}\left( y\right) }{\alpha }$ & 
$0$ & $\frac{g_{_{1}}\left( y\right) ^{2}}{\alpha ^{2}\,r^{2}}$ & $0\medskip 
$ \\ 
$0$ & $0$ & $0$ & $g_{_{1}}\left( y\right) ^{2}\medskip $%
\end{tabular}%
\right) .  \label{eq4}
\end{equation}

It should be noted that massless particles are believed to dominate the very early universe \cite{R43}. In the current methodical proposal, we consider (in section 2) massless KG-oscillators in Som-Raychaudhuri cosmic string spacetime. We study the effect rainbow gravity on the spectroscopic structure of such relativistic KG-particles. We show that the rainbow functions pair $g_{_{0}}\left( y\right)
=g_{_{1}}\left( y\right) =\left( 1-\epsilon y\right) ^{-1}$ introduces no effect, what so ever, on the spectroscopic structure of the massless KG-oscillators at hand. In subsections 2-A and 2-B, we use $g_{_{0}}\left(
y\right) =1$, $g_{_{1}}\left( y\right) =\sqrt{1-\epsilon |E|^{2}/E_{p}^{2}}$ and $g_{_{0}}\left( y\right) =1$, $g_{_{1}}\left( y\right) =\sqrt{1-\epsilon
|E|/E_{p}}$ respectively. We show that, if such rainbow functions are to serve for the invariance of the Planck's energy scale $E_{p}$, then the rainbow parameter $\epsilon $ should satisfy $\epsilon \geq 1$. However, one may switch off the rainbow parameter (i.e., set $\epsilon =0$) to retrieve the energies of KG-particles and anti-particles in Som-Raychaudhuri cosmic string pacetime without rainbow gravity. In subsection 2-C, we use $%
g_{_{0}}\left( y\right) =\left( e^{\epsilon y}-1\right) /\epsilon y$ and $%
g_{_{1}}\left( y\right) =1$ and show that the exponential structure of the rainbow function manifestly introduces a logarithmic solution for the energy levels. Consequently, such a logarithmic solution does not provide upper limits for the energies of both particles and anti-particles toward the
Planck energy scale $E_{p}$. We propose, in subsection 2-D, a new (to the best of our knowledge) rainbow functions pair, $g_{_{0}}\left( y\right)
=\left( 1-\epsilon y\right) ^{-1},\,g_{_{1}}\left( y\right) =1$. We show that this new pair reproduces the same rainbow gravity effects as those motivated by the  loop quantum gravity \cite{R41,R42}
(i.e., $g_{_{0}}\left( y\right) =1$, $g_{_{1}}\left( y\right) =\sqrt{%
1-\epsilon |E|^{n}/E_{p}^{n}};\,n=1,2$) in the sense that it preserves the invariance of the Planck's energy scale $E_{p}$ as well as symmetrization of the energy levels about $E=0$ for both KG-particles and anti-particles.  In section 3, we use our new rainbow functions pair and discuss rainbow gravity effects on massless KG-oscillators and KG-Coulombic particles in cosmic string rainbow gravity spacetime and magnetic fields. We found that the new pair shows consistent performance in the sense that it secures invariance of the Planck's energy scale $E_p$ and maintains symmetrization of the energy levels about $E=0$ value. Our concluding remarks are given in section 4.

\section{Massless KG-particles in Som-Raychaudhuri cosmic string rainbow
gravity spacetime background}

In the Som-Raychaudhuri cosmic string rainbow gravity spacetime background (\ref{eq2}), a massless KG-particle is described (in $c=\hbar =G=1$ units) by the KG-equation%
\begin{equation}
\frac{1}{\sqrt{-g}}\partial _{\mu }\sqrt{-g}g^{\mu \nu }\partial _{\upsilon
}\,\Psi =0,  \label{e7}
\end{equation}%
Which, in a straightforward manner, yields%
\begin{equation}
\left\{ -g_{_{0}}\left( y\right) ^{2}\partial _{t}^{2}+\left[ g_{_{0}}\left(
y\right) \Omega \,r\,\,\partial _{t}-\frac{g_{_{1}}\left( y\right) \,}{%
\alpha \,r}\partial _{\varphi }\right] ^{2}+\frac{g_{_{1}}\left( y\right)
^{2}}{r}\partial _{r}\,r\,\,\partial _{r}+g_{_{1}}\left( y\right)
^{2}\partial _{z}^{2}\right\} \Psi \left( t,r,\varphi ,z\right) =0.
\label{e8.1}
\end{equation}%
We may now use the substitution 
\begin{equation}
\Psi \left( t,r,\varphi ,z\right) =\exp \left( i\left[ \ell \varphi
+k_{z}z-Et\right] \right) \psi \left( r\right) =\exp \left( i\left[ \ell
\varphi +k_{z}z-Et\right] \right) \frac{R\left( r\right) }{\sqrt{r}},
\label{e8.2}
\end{equation}%
to cast our KG-equation (\ref{e8.1}) as%
\begin{equation}
\left\{ \partial _{r}^{2}-\frac{\left( \tilde{\ell}^{2}-1/4\right) }{r^{2}}-%
\tilde{\Omega}^{2}r^{2}+\tilde{\lambda}\right\} R\left( r\right) =0,
\label{e9}
\end{equation}%
where%
\begin{equation}
\tilde{\lambda}=\frac{g_{_{0}}\left( y\right) ^{2}E^{2}-2g_{_{0}}\left(
y\right) g_{_{1}}\left( y\right) \tilde{\ell}\tilde{E}-g_{_{1}}\left(
y\right) ^{2}k_{z}^{2}}{g_{_{1}}\left( y\right) ^{2}}  \label{e9.1}
\end{equation}%
and%
\begin{equation}
\tilde{\Omega}^{2}=\left( \frac{g_{_{0}}\left( y\right) }{g_{_{1}}\left(
y\right) }\tilde{E}\right) ^{2},\;\tilde{E}=\Omega E,\tilde{\ell}=\frac{\ell 
}{\alpha }.  \label{e9.2}
\end{equation}%
It is obvious that equation (\ref{e9}) resembles the two-dimensional radial
Schr\"{o}dinger oscillators, which in turn manifestly introduces the notion
of KG-oscillators. Equation (\ref{e9}) admits exact solution in the form of
hypergeometric function so that%
\begin{equation}
R\left( r\right) \sim \,r^{|\tilde{\ell}|+1/2}\exp \left( -\frac{|\tilde{%
\Omega}|\,r^{2}}{2}\right) \,\,_{1}F_{1}\left( \frac{1}{2}+\frac{|\tilde{\ell%
}|}{2}-\frac{\tilde{\lambda}}{4|\tilde{\Omega}|},1+2|\tilde{\ell}|,|\tilde{%
\Omega}|\,r^{2}\right) .  \label{e18}
\end{equation}%
However, to secure finiteness and square integrability we need to terminate the hypergeometric function into a polynomial of degree $n_{r}\geq 0$ so that the condition%
\begin{equation}
\frac{1}{2}+\frac{|\tilde{\ell}|}{2}-\frac{\tilde{\lambda}}{4|\tilde{\Omega}|%
}=-n_{r}.  \label{e19}
\end{equation}%
is satisfied. This would in turn imply that%
\begin{equation}
\tilde{\lambda}_{n_{r},\ell }=2|\tilde{\Omega}|\left( 2n_{r}+|\tilde{\ell}%
|+1\right) .  \label{e20}
\end{equation}%
and%
\begin{equation}
\psi \left( r\right) =\frac{R\left( r\right) }{\sqrt{r}}=\mathcal{N}\,r^{|%
\tilde{\ell}|}\exp \left( -\frac{|\tilde{\Omega}|\,r^{2}}{2}\right)
\,\,_{1}F_{1}\left( \frac{1}{2}+\frac{|\tilde{\ell}|}{2}-\frac{\tilde{\lambda%
}}{4|\tilde{\Omega}|},1+2|\tilde{\ell}|,|\tilde{\Omega}|\,r^{2}\right) .
\label{e201}
\end{equation}%
Consequently, Eq.(\ref{e9.1}) would read%
\begin{equation}
\left( \frac{g_{_{0}}\left( y\right) ^{2}}{g_{_{1}}\left( y\right) ^{2}}%
\right) E^{2}-2\left( \frac{g_{_{0}}\left( y\right) }{g_{_{1}}\left(
y\right) }\right) \left[ \Omega E\tilde{\ell}\,+|\Omega E|\left( 2n_{r}+|%
\tilde{\ell}|+1\right) \right] =\,k_{z}^{2}.  \label{e21}
\end{equation}

At this point, one should notice that $|\Omega E|=+\Omega _{\pm }E_{\pm }$ or $|\Omega E|=-\Omega _{\mp }E_{\pm }$ (with $\Omega _{\pm }=\pm |\Omega |$, representing positive and negative vorticities and $E_{\pm }=\pm |E|$, representing particle and anti-particle energies). This  would result%
\begin{equation}
\tilde{E}_{\pm }^{2}-2\tilde{E}_{\pm }\Omega _{\pm
}K_{+}=k_{z}^{2}\Rightarrow \tilde{E}_{\pm }=\frac{g_{_{0}}\left( y\right) }{%
g_{_{1}}\left( y\right) }E_{\pm }=\Omega _{\pm }K_{+}\pm \sqrt{\Omega
^{2}K_{+}^{2}+k_{z}^{2}},  \label{e21.1}
\end{equation}%
for $|\Omega E|=+\Omega _{\pm }E_{\pm }$, and%
\begin{equation}
\tilde{E}_{\pm }^{2}+2\tilde{E}_{\pm }\Omega _{\mp
}K_{-}=k_{z}^{2}\Rightarrow \tilde{E}_{\pm }=\frac{g_{_{0}}\left( y\right) }{%
g_{_{1}}\left( y\right) }E_{\pm }=-\Omega _{\mp }K_{-}\pm \sqrt{\Omega
^{2}K_{-}^{2}+k_{z}^{2}},  \label{e21.2}
\end{equation}%
for $|\Omega E|=-\Omega _{\mp }E_{\pm }$, where 
\begin{equation}
\tilde{E}_{\pm }=g_{_{0}}\left( y\right) E_{\pm }/g_{_{1}}\left( y\right)
,\;K_{+}=2n_{r}+|\tilde{\ell}|+\ell +1.\;K_{-}=2n_{r}+|\tilde{\ell}|-\ell +1.
\label{e21.21}
\end{equation}%
Nevertheless, the two results in (\ref{e21.1}) and (\ref{e21.2}) could be rearranged according to the vorticity signatures. That is, equation (\ref%
{e21.1}) suggests that 
\begin{equation}
\frac{g_{_{0}}\left( y\right) }{g_{_{1}}\left( y\right) }E_{+}^{\Omega
_{+}}=+|\Omega |K_{+}+\sqrt{\Omega ^{2}K_{+}^{2}+k_{z}^{2}},  \label{e21.3}
\end{equation}%
\begin{equation}
\frac{g_{_{0}}\left( y\right) }{g_{_{1}}\left( y\right) }E_{-}^{\Omega
_{-}}=-|\Omega |K_{+}-\sqrt{\Omega ^{2}K_{+}^{2}+k_{z}^{2},}  \label{e21.4}
\end{equation}%
for $|\Omega E|=+\Omega _{\pm }E_{\pm }$, and (\ref{e21.2}) yields%
\begin{equation}
\frac{g_{_{0}}\left( y\right) }{g_{_{1}}\left( y\right) }E_{+}^{\Omega
_{-}}=+|\Omega |K_{-}+\sqrt{\Omega ^{2}K_{-}^{2}+k_{z}^{2}},  \label{e21.5}
\end{equation}%
\begin{equation}
\frac{g_{_{0}}\left( y\right) }{g_{_{1}}\left( y\right) }E_{-}^{\Omega
_{+}}=-|\Omega |K_{-}-\sqrt{\Omega ^{2}K_{-}^{2}+k_{z}^{2}},  \label{e21.6}
\end{equation}%
for $|\Omega E|=-\Omega _{\mp }E_{\pm }$, where $E_{\pm }^{\Omega
_{+}}=E_{\pm }^{\left( +\right) }$ and $E_{\pm }^{\Omega _{-}}=E_{\pm
}^{\left( -\right) }$ denote energies for positive and negative vorticities, respectively. However, it would be more convenient and instructive to report these energies for positive and negative vorticities. That is,%
\begin{equation}
\frac{g_{_{0}}\left( y\right) }{g_{_{1}}\left( y\right) }E_{\pm }^{\left(
+\right) }=\pm |\Omega |K_{\pm }\pm \sqrt{\Omega ^{2}K_{\pm }^{2}+k_{z}^{2}}%
=\pm \tilde{K}_{\pm }^{\,},  \label{e22}
\end{equation}%
and 
\begin{equation}
\frac{g_{_{0}}\left( y\right) }{g_{_{1}}\left( y\right) }E_{\pm }^{\left(
-\right) }=\pm |\Omega |K_{\mp }\pm \sqrt{\Omega ^{2}K_{\mp }^{2}+k_{z}^{2}}%
=\pm \tilde{K}_{\mp }^{\,}.  \label{e23}
\end{equation}%
where%
\begin{equation}
\tilde{K}_{\pm }=\left[ |\Omega |K_{\pm }+\sqrt{\Omega ^{2}K_{\pm
}^{2}+k_{z}^{2}}\right] .  \label{e23.1}
\end{equation}

It should be noted that for all rainbow functions where $g_{_{0}}\left(
y\right) =g_{_{1}}\left( y\right) $ (including the case $g_{_{0}}\left(
y\right) =g_{_{1}}\left( y\right) =\left( 1-\epsilon y\right) ^{-1}$), equations (\ref{e22}) and (\ref{e23}) (along with (\ref{e23.1})) represent the energies of massless KG-oscillators in Som-Raychaudhuri cosmic string spacetime. The rainbow
gravity has no effect in this case. However, for $g_{_{0}}\left( y\right)
\neq g_{_{1}}\left( y\right) $ equations (\ref{e22}) and (\ref{e23})
describe the rainbow gravity effect on the spectroscopic structure of massless KG-oscillators in Som-Raychaudhuri cosmic string pacetime. Obviously, moreover, equations (\ref{e22}) and (\ref{e23}) suggest that 
\begin{equation}
E_{\pm }^{\left( -\right) }\left\vert _{\ell =\pm |\ell |}\right. =\,E_{\pm
}^{\left( +\right) }\left\vert _{\ell =\mp |\ell |}\right. ,  \label{e24}
\end{equation}%
since $K_{\pm }\left\vert _{\ell =\pm |\ell |}\right. =K_{\mp }\left\vert
_{\ell =\mp |\ell |}\right. \Rightarrow \tilde{K}_{\pm }^{\,}\left\vert
_{\ell =\pm |\ell |}\right. =\tilde{K}_{\mp }^{\,}\left\vert _{\ell =\mp
|\ell |}\right. $. It is clear that equation (\ref{e24}) identifies the so called vorticity-energy correlations. Moreover, one should be able to observe that the relation $\tilde{K}_{\pm }^{\,}\left\vert _{\ell =\pm |\ell
|}\right. =\tilde{K}_{\mp }^{\,}\left\vert _{\ell =\mp |\ell |}\right. $ identifies degeneracies associated with\ the Som-Raychaudhuri cosmic string
spacetime. These degeneracies are manifestly introduced by the fact that $%
|\Omega E|=+\Omega _{\pm }E_{\pm }$ or $|\Omega E|=-\Omega _{\mp }E_{\pm }$ (hence, such degeneracies are called spacetime associated degeneracies
(STADs)). Such STADs are observed as follows:
\begin{enumerate}
    \item  $K_{+}=\left(
2n_{r}+1\right) ;\;\forall \tilde{\ell}=-|\tilde{\ell}|$, i.e., all states with $\tilde{\ell}=-|\tilde{\ell}|\neq 0$ would merge and combine with $\ell
=0$ states for a given value of $n_{r}$, 
\item $K_{-}=\left( 2n_{r}+1\right)
;\;\forall \tilde{\ell}=+|\tilde{\ell}|$, i.e., all states with $\tilde{\ell}%
=+|\tilde{\ell}|\neq 0$ would merge and combine with $\ell =0$ states for a given value of $n_{r}$,
\item  $K_{+}=\left( 2n_{r}+2|\tilde{\ell}|+1\right)
;\;\forall \tilde{\ell}=|\tilde{\ell}|$, and 
\item $K_{-}=\left( 2n_{r}+2|%
\tilde{\ell}|+1\right) ;\;\forall \tilde{\ell}=-|\tilde{\ell}|$.
\end{enumerate}
For example, for $\forall \tilde{\ell}=+|\tilde{\ell}|$ and positive vorticity $%
\Omega =+\left\vert \Omega \right\vert $, equation (\ref{e22}) yields 
\begin{equation}
\frac{g_{_{0}}\left( y\right) }{g_{_{1}}\left( y\right) }E_{+}^{\left(
+\right) }=+|\Omega |K_{+}+\sqrt{\Omega ^{2}K_{+}^{2}+k_{z}^{2}}=|\Omega
|\left( 2n_{r}+2|\tilde{\ell}|+1\right) +\sqrt{\Omega ^{2}\left( 2n_{r}+2|%
\tilde{\ell}|+1\right) ^{2}+k_{z}^{2}},  \label{e22.1}
\end{equation}%
and%
\begin{equation}
\frac{g_{_{0}}\left( y\right) }{g_{_{1}}\left( y\right) }E_{-}^{\left(
+\right) }=-|\Omega |K_{-}+\sqrt{\Omega ^{2}K_{-}^{2}+k_{z}^{2}}=|\Omega
|\left( 2n_{r}+1\right) +\sqrt{\Omega ^{2}\left( 2n_{r}+1\right)
^{2}+k_{z}^{2}},  \label{e22.2}
\end{equation}%
whereas, equation (\ref{e23}), for $\forall \tilde{\ell}=+|\tilde{\ell}|$ and negative vorticity $\Omega =-\left\vert \Omega \right\vert $, yields%
\begin{equation}
\frac{g_{_{0}}\left( y\right) }{g_{_{1}}\left( y\right) }E_{+}^{\left(
-\right) }=|\Omega
|\left( 2n_{r}+1\right) +\sqrt{\Omega ^{2}\left( 2n_{r}+1\right)
^{2}+k_{z}^{2}},  \label{e23.2}
\end{equation}%
and%
\begin{equation}
\frac{g_{_{0}}\left( y\right) }{g_{_{1}}\left( y\right) }E_{-}^{\left(
-\right) }=|\Omega
|\left( 2n_{r}+2|\tilde{\ell}|+1\right) +\sqrt{\Omega ^{2}\left( 2n_{r}+2|%
\tilde{\ell}|+1\right) ^{2}+k_{z}^{2}}.  \label{e23.3}
\end{equation}

At this point, it should be made clear that the above mentioned vorticity-energy correlations as well as STADs are specifically consequences/byproducts of the Som-Raychaudhuri cosmic string spacetime and have nothings to do with rainbow gravity. However, it is obvious that rainbow gravity will affect the energies of the probe massless KG-oscillators for $g_{_{0}}\left( y\right)
\neq g_{_{1}}\left( y\right) $. In what follows we report the rainbow gravity effects for different fine-tuned rainbow functions used in the literature.

\subsection{Rainbow functions $g_{_{0}}\left( y\right) =1$, $%
g_{_{1}}\left( y\right) =\sqrt{1-\epsilon E^{2}/E_{p}^{2}}$}

For the rainbow functions $g_{_{0}}\left( y\right) =1$, $g_{_{1}}\left(
y\right) =\sqrt{1-\delta E^{2}};\;\delta =\epsilon /E_{p}^{2}$, we re-cast (%
\ref{e22}) and (\ref{e23}) as%
\begin{equation}
\frac{E_{\pm }^{\left( +\right) \,2}}{1-\delta E_{\pm }^{\left( +\right) \,2}%
}=\tilde{K}_{\pm }^{2}\Rightarrow E_{\pm }^{\left( +\right) }=\pm \frac{%
\tilde{K}_{\pm }}{\sqrt{\tilde{K}_{\pm }^{2}\delta +1}}=\pm \frac{E_{p}}{%
\sqrt{\epsilon +\left( \frac{E_{p}}{\tilde{K}_{\pm }^{\left( +\right) \,}}%
\right) ^{2}}},  \label{e25}
\end{equation}%
and \ 
\begin{equation}
\frac{E_{\pm }^{\left( -\right) \,2}}{1-\delta E_{\pm }^{\left( -\right) \,2}%
}=\tilde{K}_{\mp }^{\,2}\Rightarrow E_{\pm }^{\left( -\right) }=\pm \frac{%
\tilde{K}_{\mp }^{\,}}{\sqrt{\tilde{K}_{\mp }^{2}\delta +1}}=\pm \frac{E_{p}%
}{\sqrt{\epsilon +\left( \frac{E_{p}}{\tilde{K}_{\pm }^{\left( -\right) \,}}%
\right) ^{2}}},  \label{e26}
\end{equation}%
Notably, when expanding about $\delta =0$, we get $\left\vert E_{\pm
}^{\left( +\right) }\right\vert \sim \tilde{K}_{\pm }-\tilde{K}_{\pm
}^{3}\delta /2+O\left( \delta ^{2}\right) $ and $\left\vert E_{\pm }^{\left(
-\right) }\right\vert \sim \tilde{K}_{\mp }^{\,}-\tilde{K}_{\mp }^{2}\delta
/2+O\left( \delta ^{2}\right) $. That is, the energies in both cases are less than the energies when rainbow gravity is switched off (i.e., at $%
\delta =0$ we have $\left\vert E_{\pm }^{\left( +\right) }\right\vert =%
\tilde{K}_{\pm }$ and $\left\vert E_{\pm }^{\left( -\right) }\right\vert =%
\tilde{K}_{\mp }^{\,}$). Yet, for $|\Omega |\rightarrow
\infty $ we have $\tilde{K}_{\pm }^{\,}\rightarrow 2|\Omega |K_{\pm }$ so that a Taylor expansion would yield 
\begin{equation}
E_{\pm }^{\left( \pm \right) }\simeq \pm 1/\sqrt{\delta }+O\left( 1/\Omega
^{2}\right) = \pm E_{p}/\sqrt{\epsilon }\Rightarrow |E_{\pm }^{\left(
+\right) }|= E_{p}/\sqrt{\epsilon }.  \label{e26.1}
\end{equation}%
This would suggest that, as long as rainbow gravity model is in point, the rainbow parameter $\epsilon $ should satisfy $\epsilon \geq 1$. Which would, in turn, emphasis/document that the Planck energy $E_{p}$ is the maximum possible probe particle and anti-particle energy.
\begin{figure}[!ht]  
\centering
\includegraphics[width=0.3\textwidth]{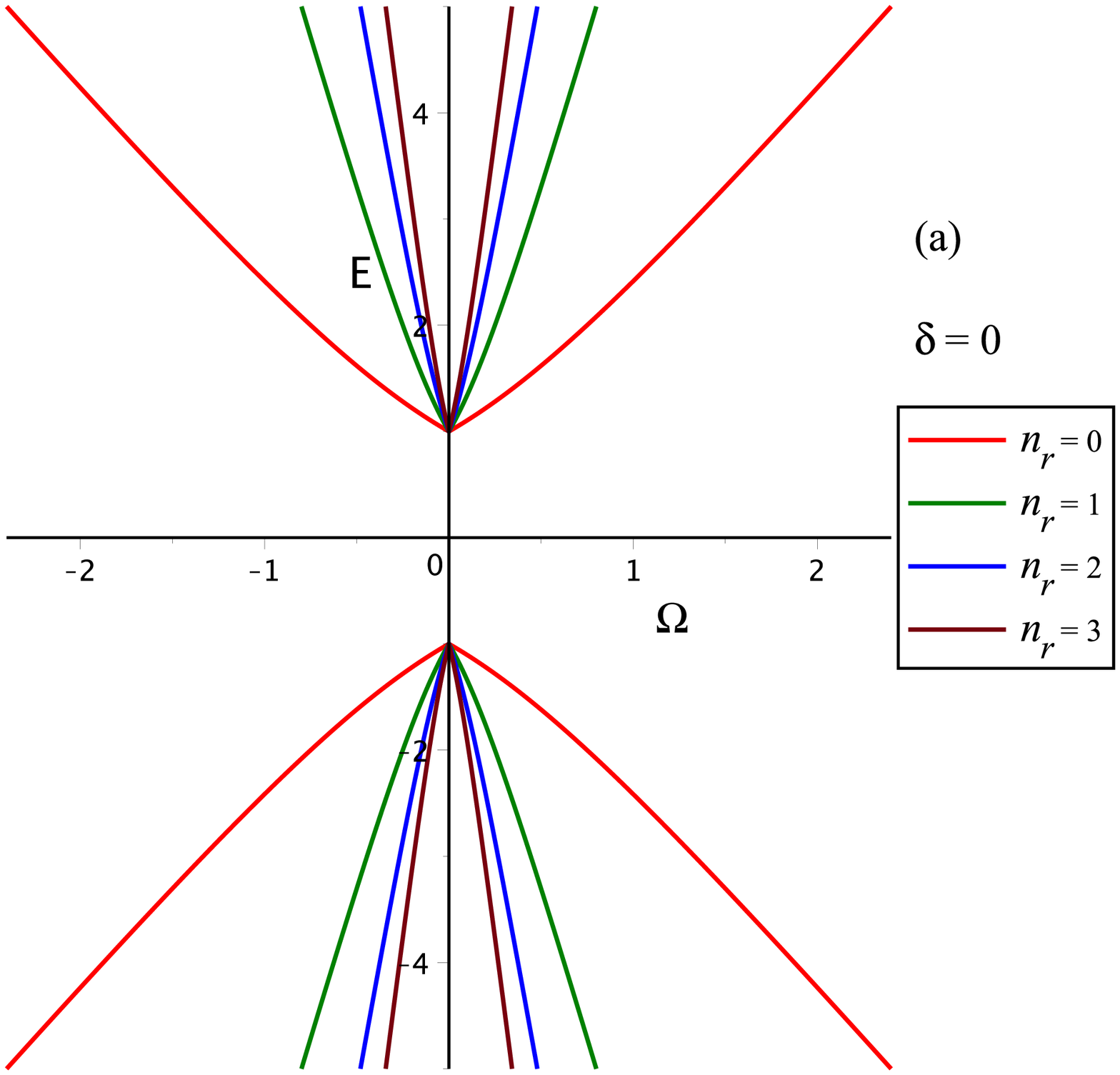}
\includegraphics[width=0.3\textwidth]{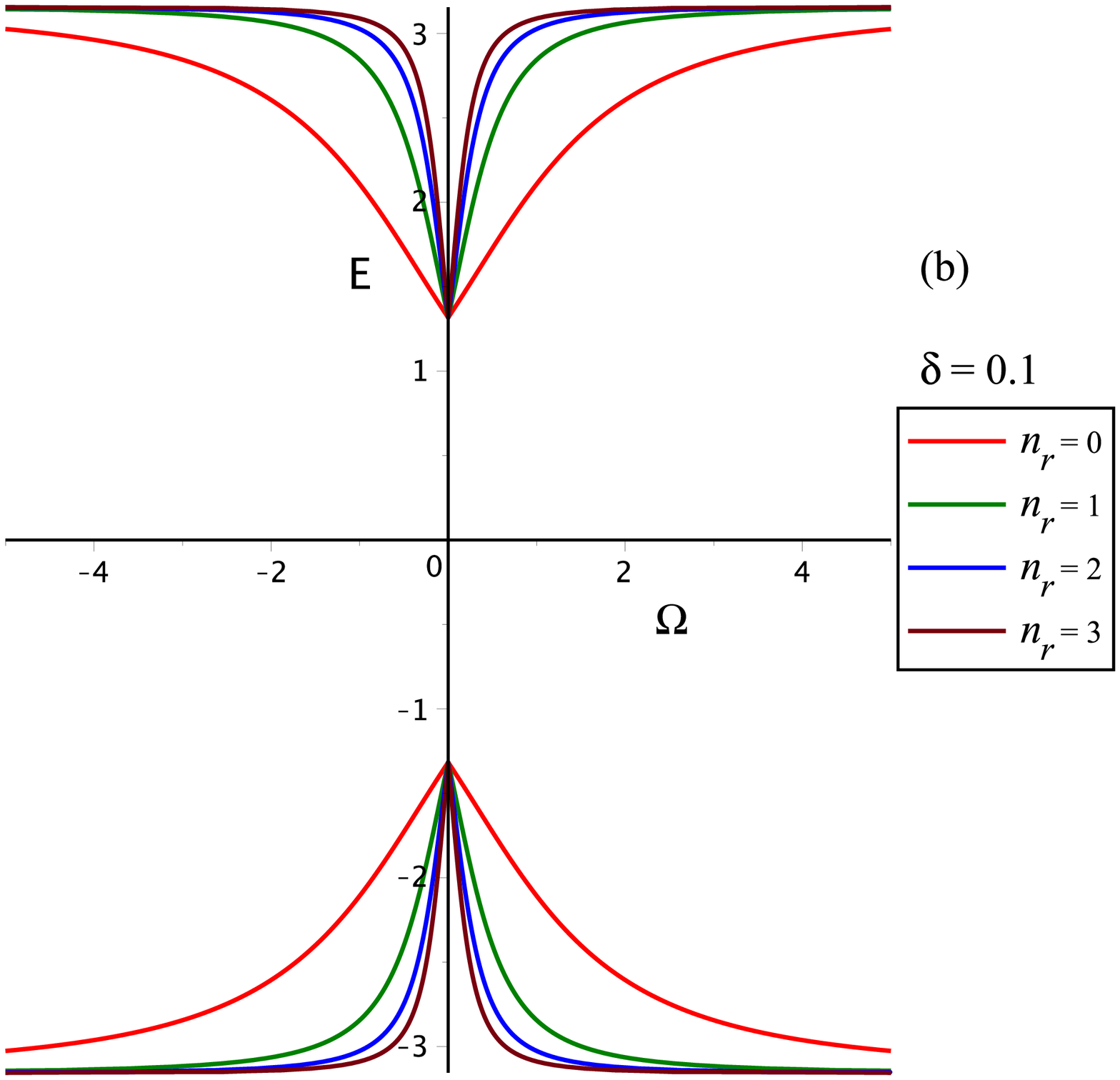} 
\includegraphics[width=0.3\textwidth]{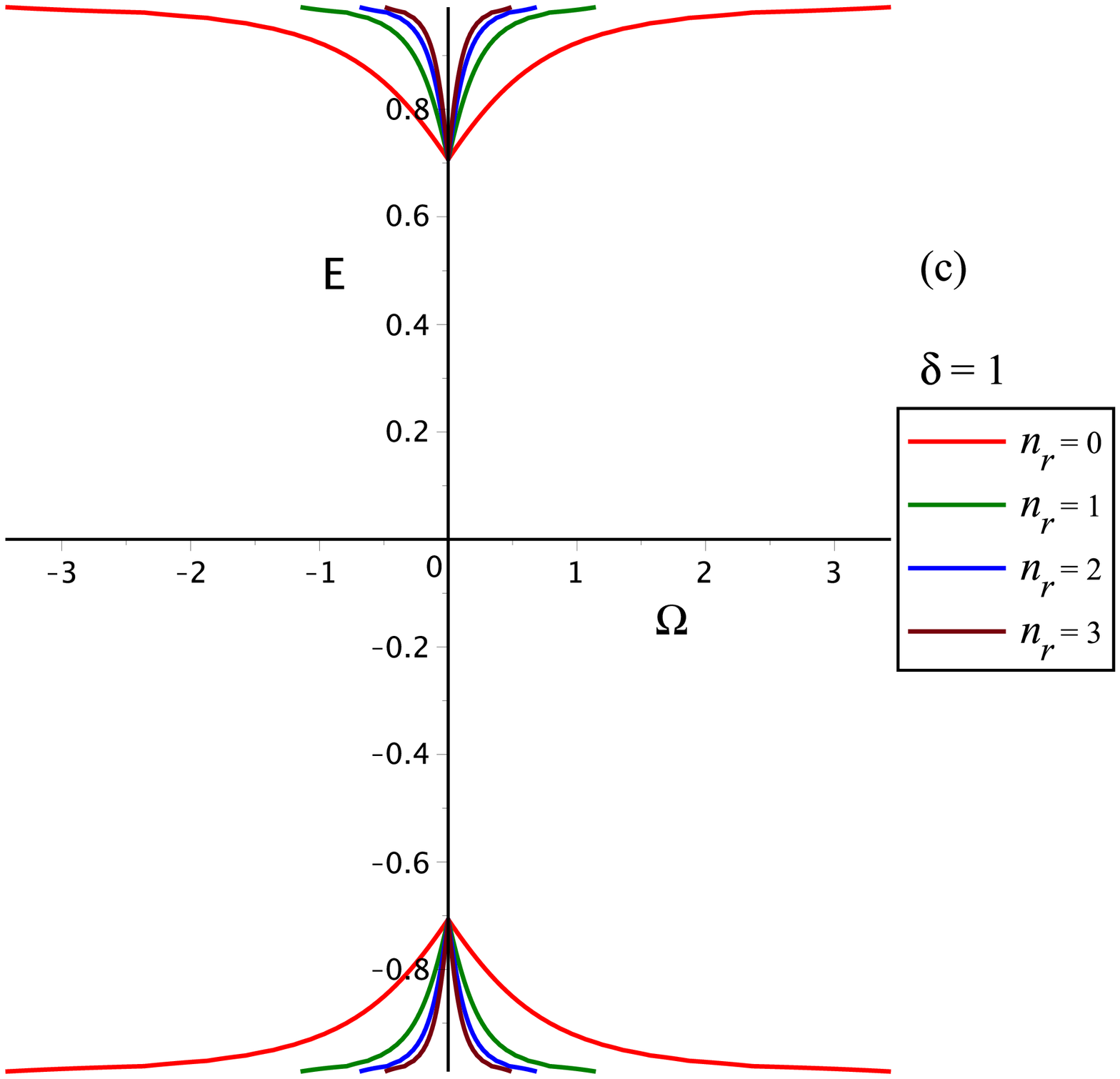}
\caption{\small 
{ The energy levels of (\ref{e25}) and (\ref{e26}) at
different vorticity parameter $\Omega $ values and for $g_{_{0}}\left(
y\right) =1$, $g_{_{1}}\left( y\right) =\sqrt{1-\delta E^{2}}$, using $%
\alpha =0.5$, $k_{z}=1$, $\ell =0$, and $n_{r}=0,1,2,3$ so that (a) for $%
\delta =0$, i.e., no rainbow gravity effects, $\epsilon =0$. (b) for $\delta
=0.1$, and (c) for $\delta =1$.}}
\label{fig1}
\end{figure}%
\begin{figure}[!ht]  
\centering
\includegraphics[width=0.3\textwidth]{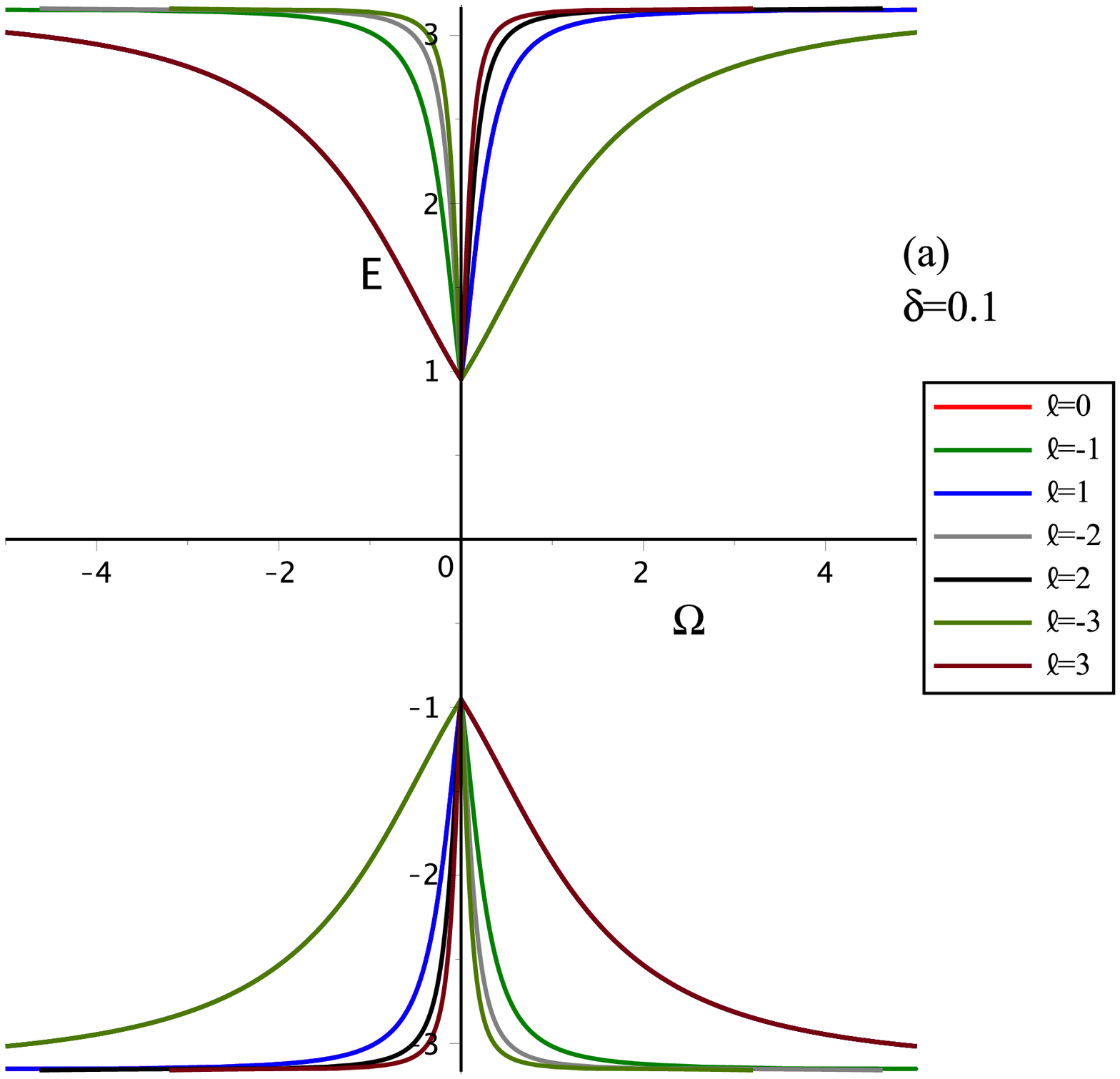}
\includegraphics[width=0.3\textwidth]{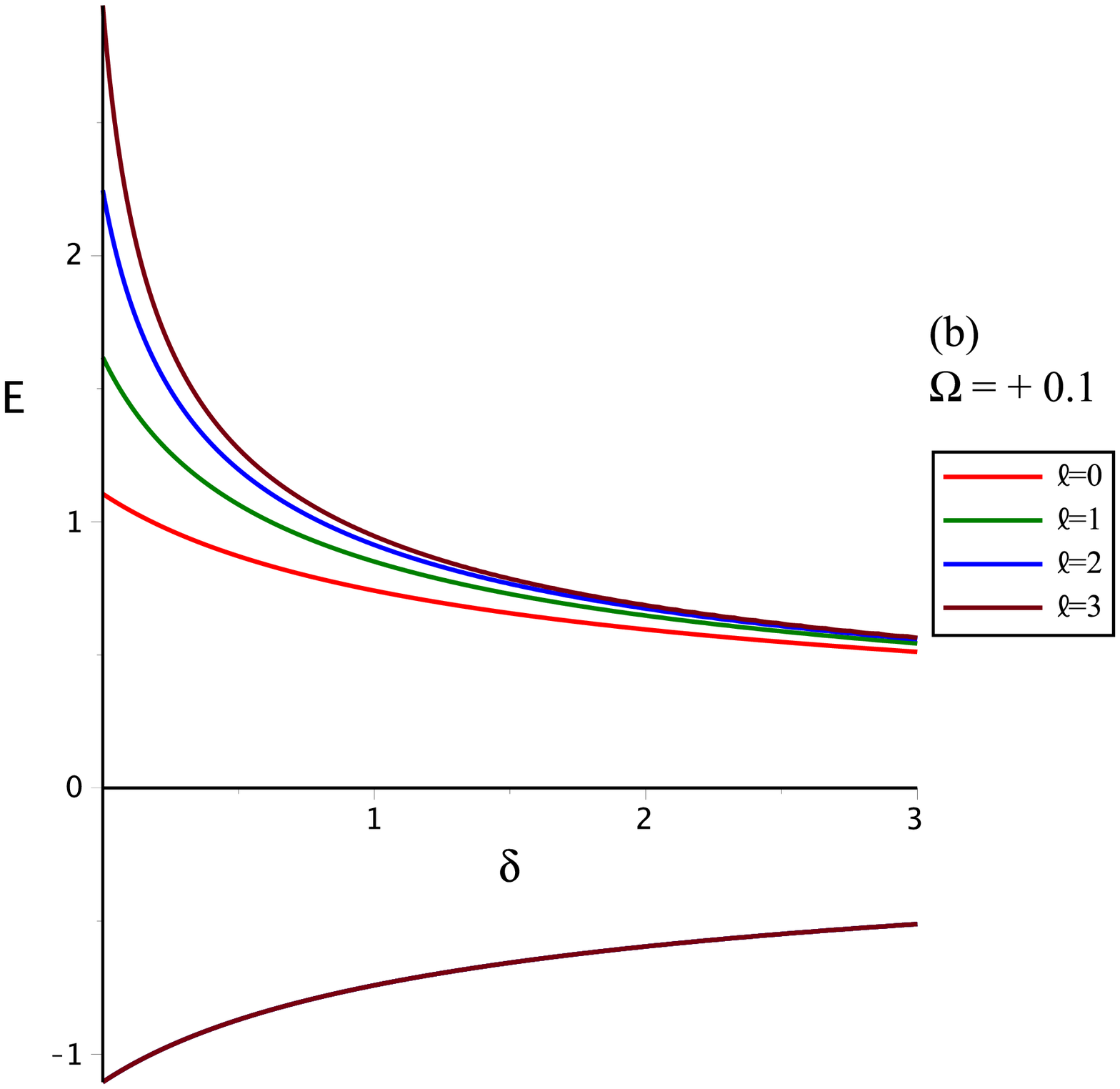} 
\includegraphics[width=0.3\textwidth]{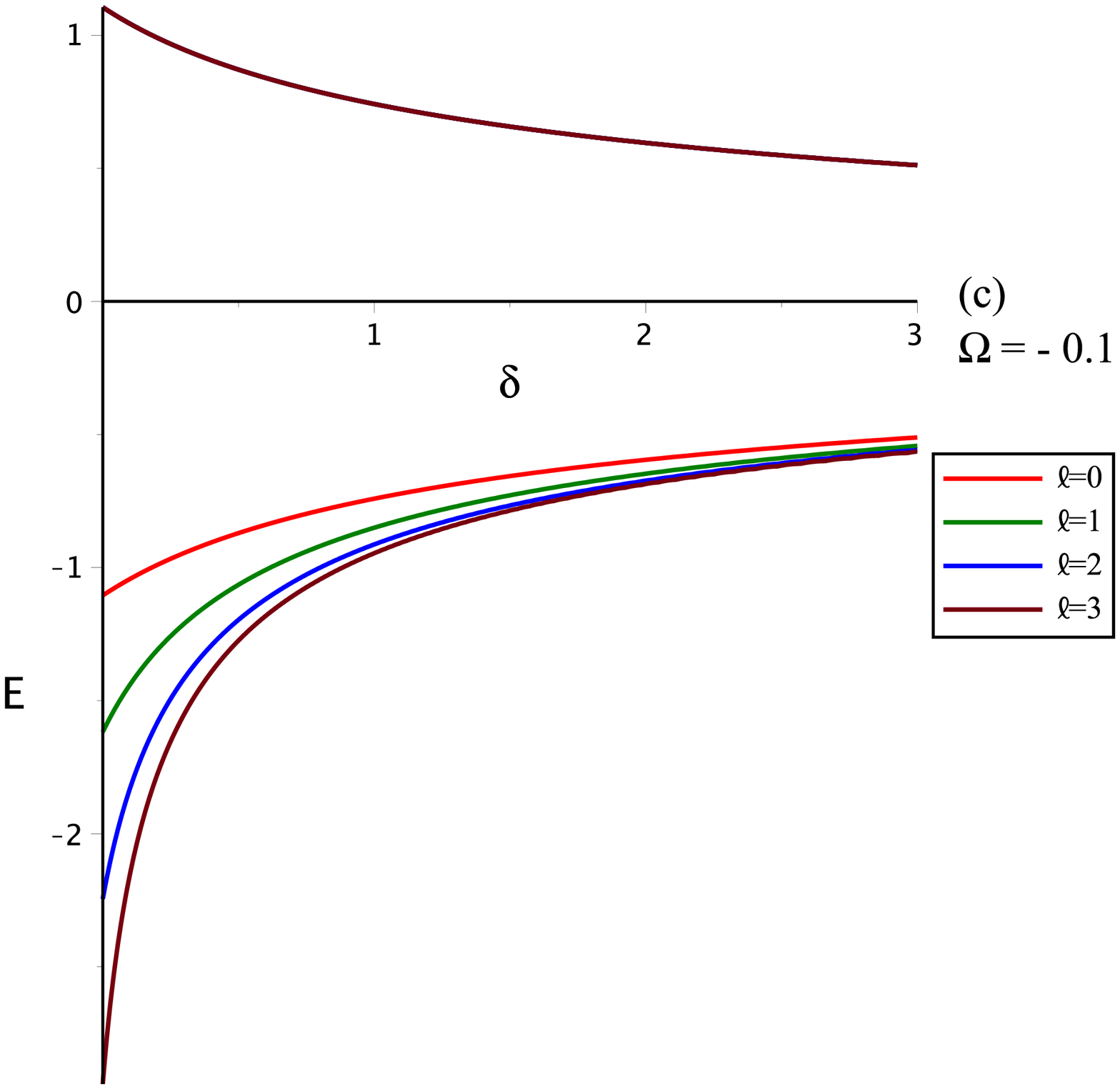}
\caption{\small 
{ The energy levels of (\ref{e25}) and (\ref{e26}) for $%
g_{_{0}}\left( y\right) =1$, $g_{_{1}}\left( y\right) =\sqrt{1-\delta E^{2}}$%
, using $\alpha =0.5$, $k_{z}=1$, and $n_{r}=0$ so that (a) shows $E$
against $\Omega $ for $\delta =0.1$ and $\ell =0,\pm 1,\pm 2,\pm 3$, (b)
shows $E$ against $\delta $ for $\Omega =+0.1$ and $\ell =0,1,2,3$, and (c)
shows $E$ against $\delta $ for $\Omega =-0.1$ and $\ell =0,1,2,3$.}}
\label{fig2}
\end{figure}%

To observe the effect of rainbow gravity, we plot in Figure 1(a) and 1(b) the energy levels without ($\delta=0$) and with rainbow gravity ($\delta =0.1$), respectively, for $n_{r}=0,1,2,3$ and $\ell =0$ (i.e., $s$-states) at different values for the vorticity parameter $\Omega $. Moreover, to observe the effect of the rainbow parameter $\delta $, we plot in Figure 1(c) the same energy levels for $\delta =1$. A comparison between Figures 1(a) and 1(b) clearly documents that rainbow gravity puts an upper bound for the energies $\left\vert E_{\pm }^{\left( \pm \right) }\right\vert _{\max .}=1/%
\sqrt{\delta }=1/\sqrt{0.1}\simeq 3.16$, as given 
in (\ref{e26.1}), A comparison between Figures 1(b), at $\delta =0.1$, and 1(c), at $\delta =1$, shows that the energy levels are pushed closer to $E=0$ value and the energy gap about $E=0$ gets narrower for larger $\delta $ values. It is also obvious that $s$-states, $\ell =0$, are symmetric about $E=0$ value (as shown in 1(a), 1(b) and 1(c)). This symmetry breaks when different magnetic quantum numbers $\ell \neq 0$ are considered. This is documented in Figures 2(a), 2(b), and 2(c). Figure 2(a) shows the energies for $n_{r}=0$, $\ell
=0,\pm 1,\pm 2,\pm 3$, at $\delta =0.1$ and different values for vorticity parameter $\Omega $. One observes that the spacetime associated degeneracies (STADs) discussed above become active and break the symmetry of the energy levels about $E=0$ value. To clearly show how STADs (discussed for (\ref{e24})) work, we plot in 2(b) and 2(c) for $\Omega =+0.1$ and $\Omega =-0.1$, respectively, using $n_{r}=0$, $\ell =0,1,2,3$ for different $\delta $ values. That is, for $\Omega =+0.1$ we have to use (\ref{e25}), where $%
E_{+}^{(+)}=+\tilde{K}_{+}/\sqrt{\tilde{K}_{+}^{2}\delta +1}$, $K_{+}=\left(
2n_{r}+2|\tilde{\ell}|+1\right) ;\;\forall \tilde{\ell}=|\tilde{\ell}|$, and $E_{-}^{(+)}=-\tilde{K}_{-}/\sqrt{\tilde{K}_{-}^{2}\delta +1}$, $%
K_{+}=\left( 2n_{r}+1\right) ;\;\forall \tilde{\ell}=|\tilde{\ell}|$ (which is an obvious observation of Fig.2(b)). Whereas, for $\Omega =-0.1$ we have to use (\ref{e26}), where $E_{+}^{(-)}=+\tilde{K}_{-}/\sqrt{\tilde{K}%
_{-}^{2}\delta +1}$, and $E_{-}^{(-)}=-\tilde{K}_{+}/\sqrt{\tilde{K}%
_{+}^{2}\delta +1}$(as obvious in Fig.2(c)).

\subsection{Rainbow functions $g_{_{0}}\left( y\right) =1$, $%
g_{_{1}}\left( y\right) =\sqrt{1-\epsilon \left\vert
E\right\vert /E_{p}}$}

With the rainbow functions $g_{_{0}}\left( y\right) =1$, $g_{_{1}}\left(
y\right) =\sqrt{1-2\beta \left\vert E\right\vert }$, where $\,\beta
=\epsilon /2E_{p}$ and $\left\vert E\right\vert =\pm E_{\pm }$, equations (\ref{e22}) and (\ref{e23}), respectively, imply that%
\begin{equation}
\frac{E_{\pm }^{\left( +\right) \,2}}{1-2\beta \left\vert E_{\pm }^{\left(
+\right) \,}\right\vert }=\tilde{K}_{\pm }^{\,2}\Rightarrow E_{\pm }^{\left(
+\right) \,2}+2\beta \left\vert E_{\pm }^{\left( +\right) \,}\right\vert =%
\tilde{K}_{\pm }^{\,2},\text{ }  \label{e27}
\end{equation}%
and%
\begin{equation}
\frac{E_{\pm }^{\left( -\right) \,2}}{1-2\beta |E_{\pm }^{\left( -\right)
\,}|}=\tilde{K}_{\mp }^{2}\Rightarrow E_{\pm }^{\left( -\right) \,2}+2\beta
|E_{\pm }^{\left( -\right) \,}|=\tilde{K}_{\mp }^{2}.  \label{e28}
\end{equation}
\begin{figure}[!ht]  
\centering
\includegraphics[width=0.3\textwidth]{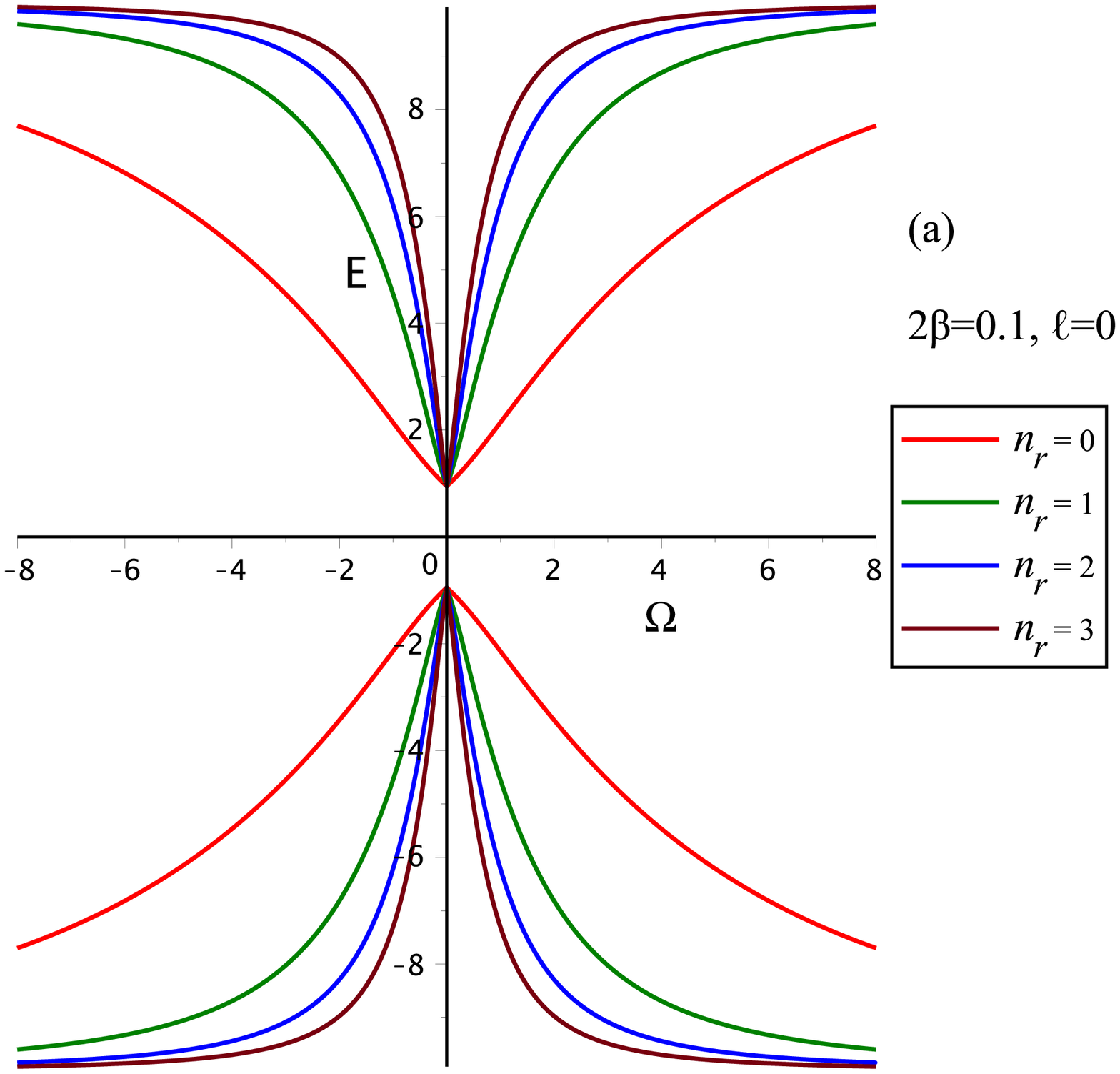}
\includegraphics[width=0.3\textwidth]{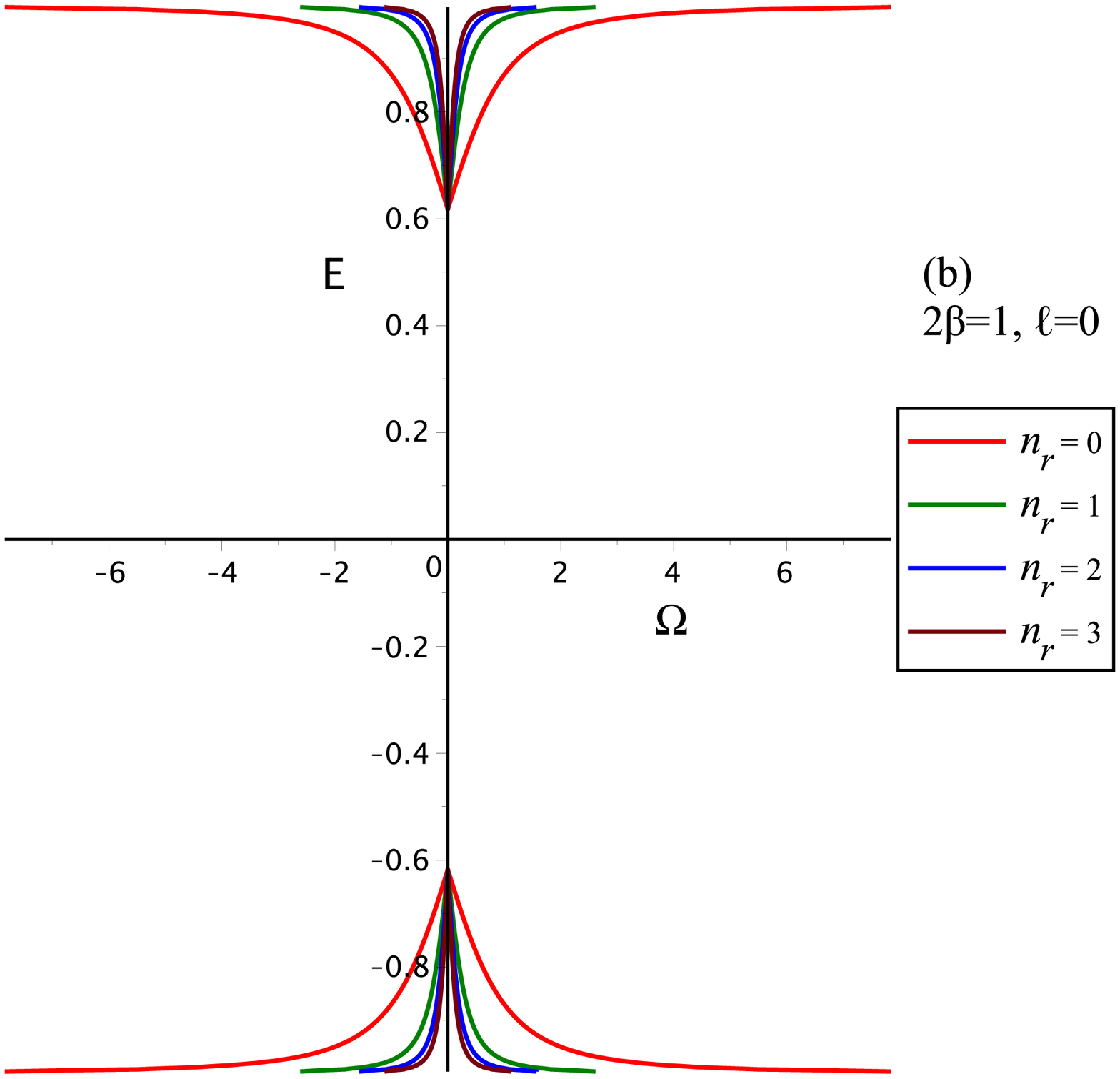} 
\includegraphics[width=0.3\textwidth]{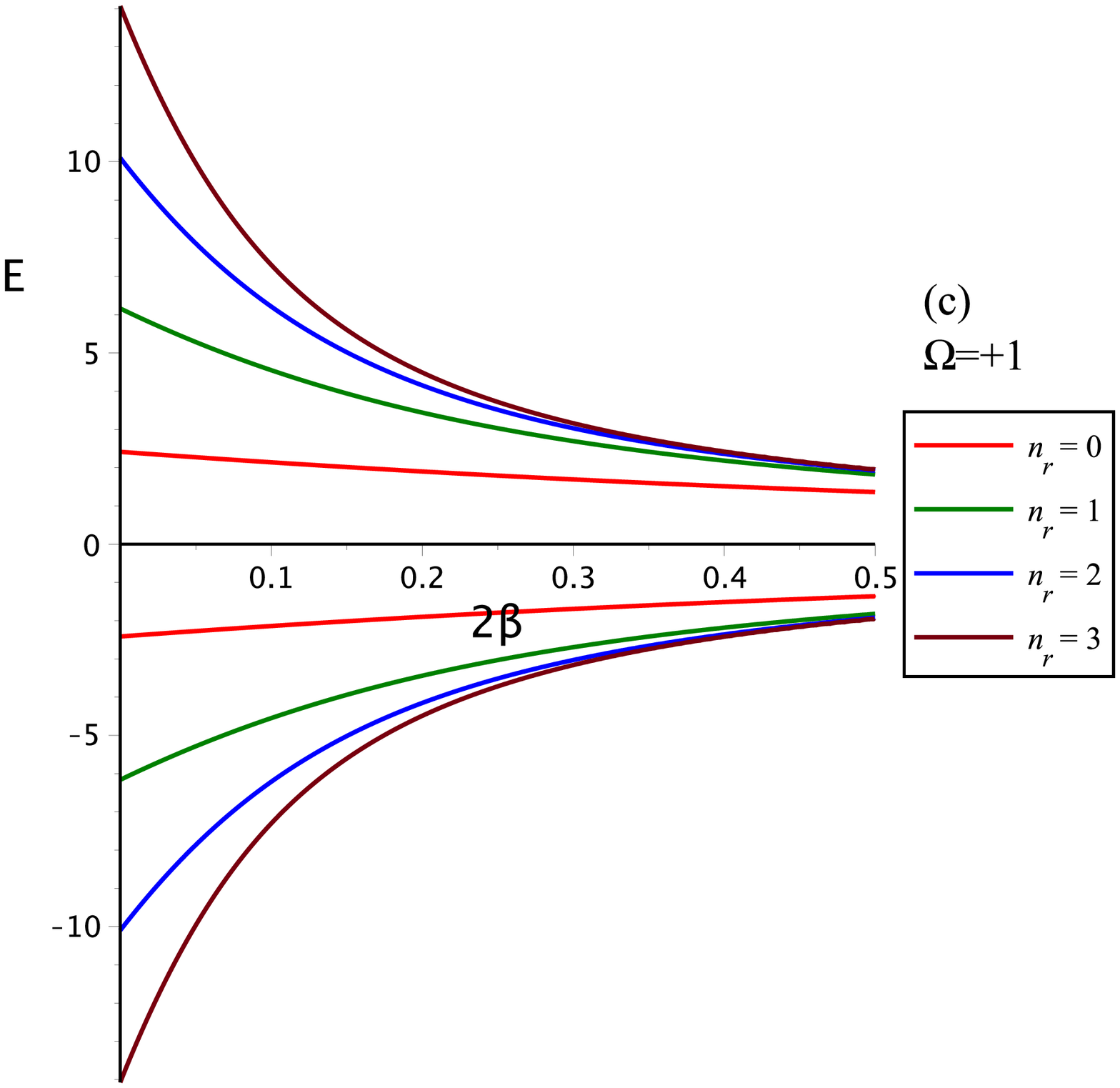}
\caption{\small 
{ The energy levels of (\ref{e27}) and (\ref{e28}) for $%
g_{_{0}}\left( y\right) =1$, $g_{_{1}}\left( y\right) =\sqrt{1-\beta
\left\vert E\right\vert }$, using $\alpha =0.5$, $k_{z}=1$, $\ell =0$, and $%
n_{r}=0,1,2,3$ so that (a) shows $E$ against $\Omega $ for $\beta =0.1$, (b)
shows $E$ against $\Omega $ for $\beta =1$, and (c) shows $E$ against $\beta 
$ for $\Omega =+1$.}}
\label{fig3}
\end{figure}%

Under such settings, one would use (\ref{e27}), with $\left\vert
E\right\vert =\pm E_{\pm }^{\left( +\right) }=\pm E_{\pm }^{\left( -\right)
} $, to obtain%
\begin{equation}
\left\{ 
\begin{tabular}{l}
$\left( E_{+}^{\left( +\right) \,}+\beta \tilde{K}_{\pm }^{\,2}\right) ^{2}=%
\tilde{K}_{\pm }^{\,2}+\beta ^{2}\tilde{K}_{\pm }^{\,4}$ \\ 
$\left( E_{-}^{\left( +\right) \,}-\beta \tilde{K}_{\pm }^{\,2}\right) ^{2}=%
\tilde{K}_{\pm }^{\,2}+\beta ^{2}\tilde{K}_{\pm }^{\,4}$%
\end{tabular}%
\right\} \Longrightarrow E_{\pm }^{\left( +\right) \,}=\mp \beta \tilde{K}%
_{\pm }^{\,2}\pm \sqrt{\tilde{K}_{\pm }^{\,2}+\beta ^{2}\tilde{K}_{\pm
}^{\,4}}  \label{e27-1}
\end{equation}%
and use (\ref{e28}) to obtain%
\begin{equation}
\left\{ 
\begin{tabular}{l}
$\left( E_{+}^{\left( -\right) \,}+\beta \tilde{K}_{\mp }^{\,2}\right) ^{2}=%
\tilde{K}_{\mp }^{\,2}+\beta ^{2}\tilde{K}_{\mp }^{\,4}$ \\ 
$\left( E_{-}^{\left( -\right) \,}-\beta \tilde{K}_{\mp }^{\,2}\right) ^{2}=%
\tilde{K}_{\mp }^{\,2}+\beta ^{2}\tilde{K}_{\mp }^{\,4}$%
\end{tabular}%
\right\} \Longrightarrow E_{\pm }^{\left( -\right) \,}=\mp \beta \tilde{K}%
_{\mp }^{\,2}\pm \sqrt{\tilde{K}_{\mp }^{\,2}+\beta ^{2}\tilde{K}_{\mp
}^{\,4}}  \label{e28-1}
\end{equation}%
Clearly, the symmetry about $E_{\pm }^{\left( \pm \right) }=0$ value is unbroken as a result of the structure of the rainbow functions. This may clearly be observed in Figures 3(a), 3(b), and 3(c). Moreover, for $|\Omega |\rightarrow \infty $ a Taylor expansion would yield 
\begin{equation}
E_{+}^{\left( +\right) }\left\vert _{|\Omega |=\infty }\right. \simeq \frac{1%
}{2\beta }+O\left( \frac{1}{\Omega ^{2}}\right) \simeq \frac{E_{p}}{\epsilon 
},\text{ and }\;E_{-}^{\left( +\right) }\left\vert _{|\Omega |=\infty
}\right. \sim -\frac{1}{2\beta }+O\left( \frac{1}{\Omega ^{2}}\right)
\label{e27.1}
\end{equation}%
for (\ref{e27}). Similarly,%
\begin{equation}
E_{+}^{\left( -\right) }\left\vert _{|\Omega |=\infty }\right. \simeq \frac{1%
}{2\beta }+O\left( \frac{1}{\Omega ^{2}}\right) \simeq \frac{E_{p}}{\epsilon 
},\;\text{and }E_{-}^{\left( -\right) }\left\vert _{|\Omega |=\infty
}\right. \sim -\frac{1}{2\beta }+O\left( \frac{1}{\Omega ^{2}}\right)
\label{e28.1}
\end{equation}%
for (\ref{e28}). In Figure 3(a) our $2\beta=0.1\Rightarrow |E_{\pm}^{(\pm)}|_{max}=10$ and in 4(b) our $2\beta=1\Rightarrow |E_{\pm}^{(\pm)}|_{max}=1$ are obvious energy bounds. Hence our probe KG-particles' and anti-particles' energies comply with the DSR/rainbow gravity model in such a way that $%
\left\vert E_{\pm }^{\left( \pm \right) }\right\vert \leq E_{p}$, provided that the rainbow parameter satisfies the relation $\epsilon \geq 1$.  

\subsection{Rainbow functions $g_{_{0}}\left( y\right) =\left( e^{%
\epsilon y}-1\right) /\epsilon y$ and $g_{_{1}}\left(
y\right) =1$, $y=\left\vert E\right\vert /E_{p}$}

Under such rainbow functions setting, equations (\ref{e22}) and (\ref{e23})
would, respectively, yield%
\begin{equation}
\left( \frac{e^{\zeta \left\vert E\right\vert }-1}{\zeta \left\vert
E\right\vert }\right) E_{\pm }^{\left( +\right) }=\pm \tilde{K}_{\pm
}^{\,}\Rightarrow \left( e^{\zeta \left\vert E\right\vert }-1\right) E_{\pm
}^{\left( +\right) }=\pm \zeta \left\vert E\right\vert \tilde{K}_{\pm
}^{\,}\Rightarrow \left( e^{\pm \zeta E_{\pm }^{\left( +\right) }}-1\right)
=\zeta \tilde{K}_{\pm }^{\,},  \label{e29}
\end{equation}%
\begin{figure}[!ht]  
\centering
\includegraphics[width=0.35\textwidth]{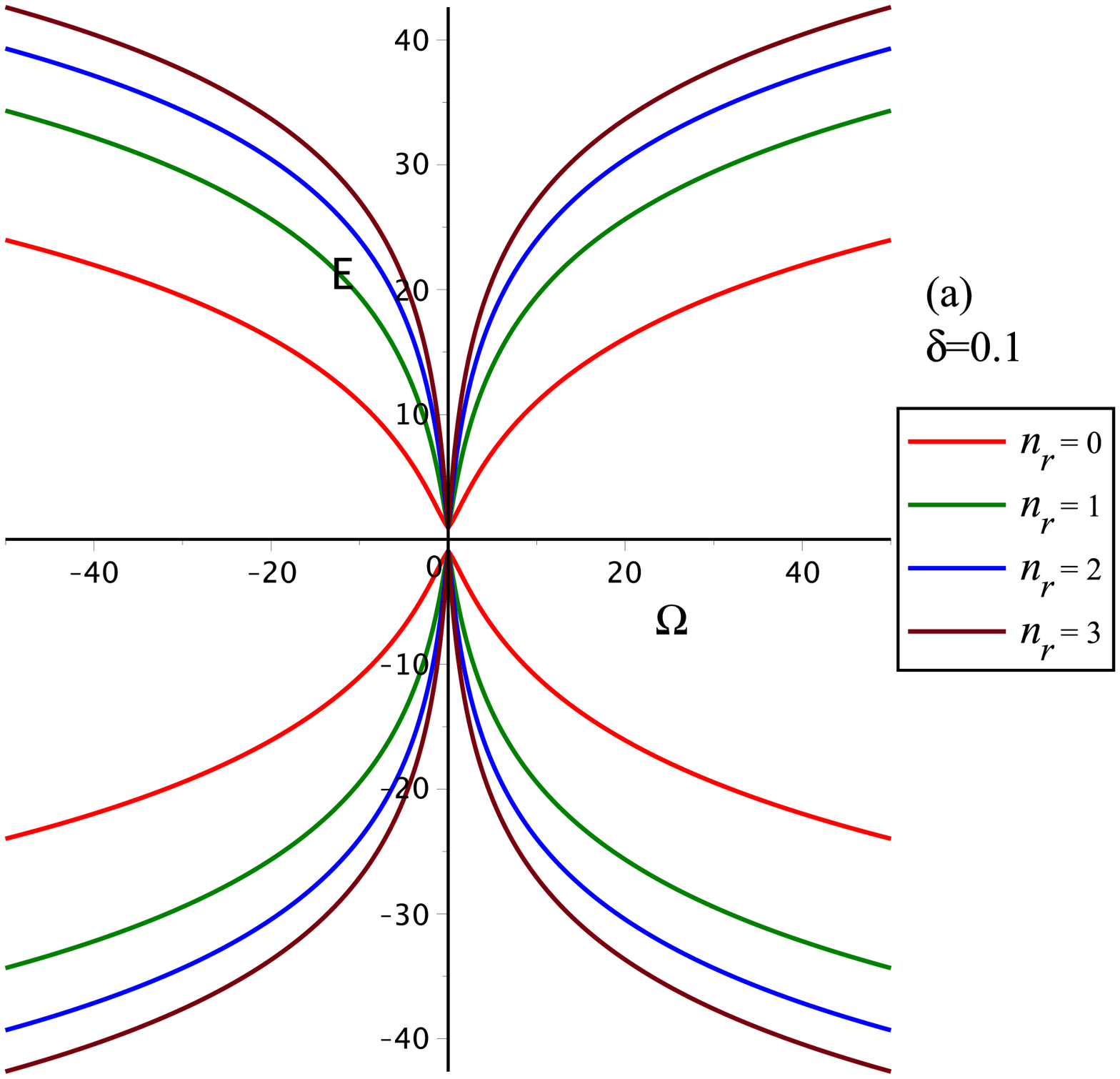}
\includegraphics[width=0.35\textwidth]{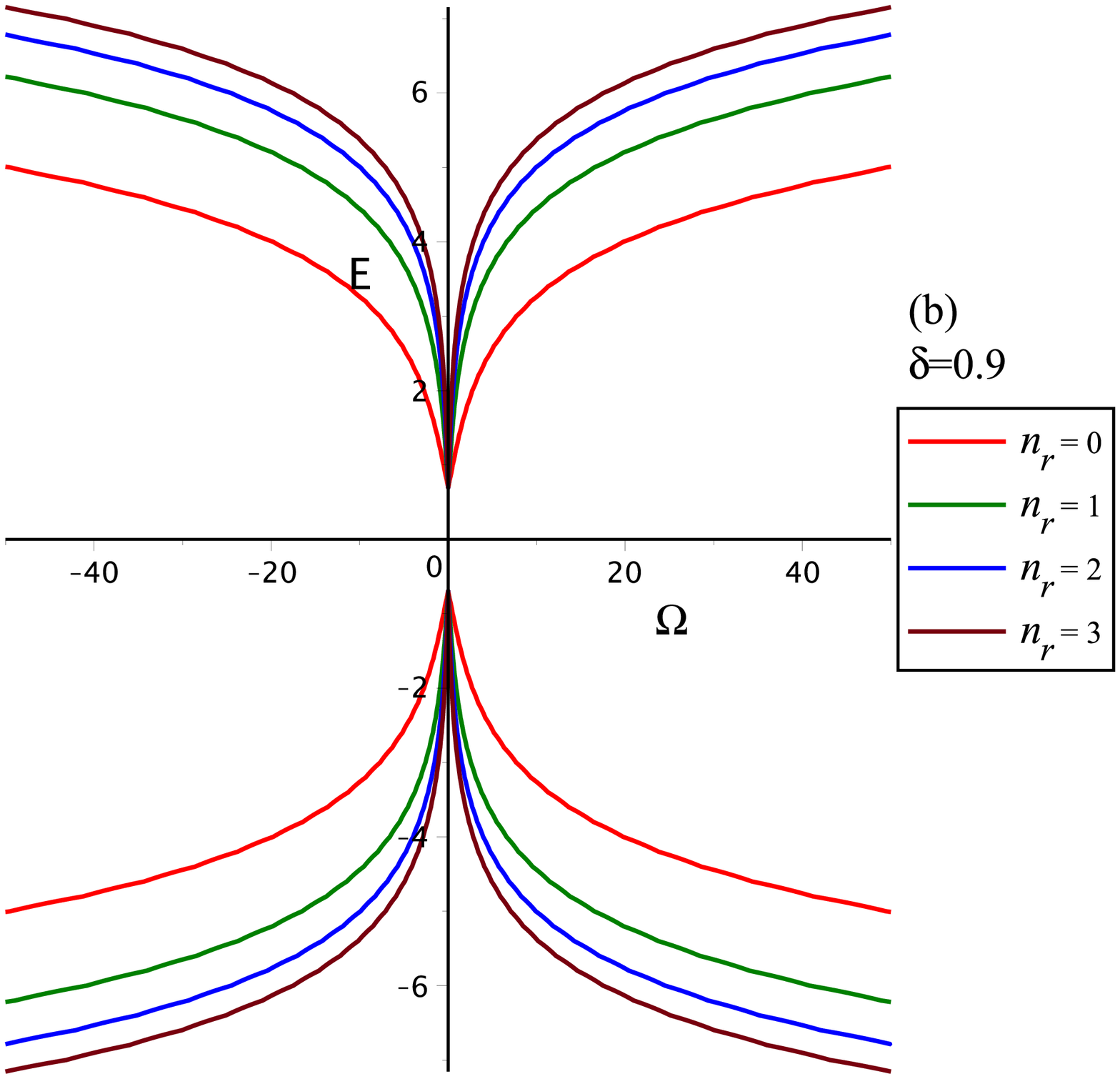} 
\caption{\small 
{ The energy levels of (\ref{e31}) and (\ref{e32}) for\textbf{%
\ }$g_{_{0}}\left( y\right) =\left( e^{\zeta \left\vert E\right\vert
}-1\right) /\zeta \left\vert E\right\vert $, $g_{_{1}}\left( y\right) =1$,
using $\alpha =0.5$, $k_{z}=1$, $\ell =0$, and $n_{r}=0,1,2,3$ so that (a)
shows $E$ against $\Omega $ for $\zeta =0.1$, and (b) shows $E$ against $%
\Omega $ for $\zeta =0.9$.}}
\label{fig4}
\end{figure}%
and%
\begin{equation}
\left( \frac{e^{\zeta \left\vert E\right\vert }-1}{\zeta \left\vert
E\right\vert }\right) E_{\pm }^{\left( -\right) }=\pm \tilde{K}_{\mp
}^{\,}\Rightarrow \left( e^{\zeta \left\vert E\right\vert }-1\right) E_{\pm
}^{\left( -\right) }=\pm \zeta \left\vert E\right\vert \tilde{K}_{\mp
}^{\,}\Rightarrow \left( e^{\pm \zeta E_{\pm }^{\left( -\right) }}-1\right)
=\zeta \tilde{K}_{\mp }^{\,},  \label{e30}
\end{equation}%
where $\zeta =\epsilon /E_{p}$. Consequently, (\ref{e29}), with $\left\vert
E\right\vert =\pm E_{\pm }^{\left( +\right) }=\pm E_{\pm }^{\left( -\right)
} $, would read%
\begin{equation}
\left( e^{\pm \zeta E_{\pm }^{\left( +\right) }}-1\right) =\zeta \tilde{K}%
_{\pm }^{\,}\Rightarrow E_{\pm }^{\left( +\right) }=\pm \frac{1}{\zeta }\ln
\left( 1+\zeta \tilde{K}_{\pm }^{\,}\right) ,  \label{e31}
\end{equation}%
and (\ref{e30}) reads%
\begin{equation}
\left( e^{\pm \zeta E_{\pm }^{\left( -\right) }}-1\right) =\zeta \tilde{K}%
_{\mp }^{\,}\Rightarrow E_{\pm }^{\left( -\right) }=\pm \frac{1}{\zeta }\ln
\left( 1+\zeta \tilde{K}_{\mp }^{\,}\right)  \label{e32}
\end{equation}%
It is obvious one observes that such energy levels are unbounded and may grow indefinitely as a result of the logarithmic term ( which grows up with increasing frequency $\left\vert \Omega \right\vert $, in (\ref{e23.1}), of the KG-oscillator at hand), As such, one should not expect that the corresponding energies $\left\vert E_{\pm }^{\left( \pm \right) }\right\vert 
$ would be less than the Planck energy $E_{p}$. Such a rainbow functions structure violates the Planck energy scale invariance, therefore. 

However, the rainbow function $g_{_{0}}\left( y\right) =\left( e^{\epsilon
y}-1\right) /\epsilon y\rightarrow 1$ as $\epsilon \rightarrow 0$. Consequently, $E_{\pm }^{\left( +\right) }=\pm \tilde{K}_{\pm }^{\,}$ and $%
E_{\pm }^{\left( -\right) }=\pm \tilde{K}_{\mp }^{\,}$ when rainbow gravity is switched off (i.e., for KG-oscillators in Som-Raychaudhuri cosmic string spacetime, without the effect of rainbow gravity). Nevertheless, with the effect of rainbow gravity, a Taylor expansion around $\zeta =0$ for (\ref%
{e31}) and (\ref{e32}) would, respectively, imply%
\begin{equation}
E_{\pm }^{\left( +\right) }\simeq \pm \left[ \tilde{K}_{\pm }^{\,}-\frac{1}{2%
}\zeta \tilde{K}_{\pm }^{\,2}+O\left( \zeta ^{2}\right) \right] \Rightarrow
\left\vert E_{\pm }^{\left( +\right) }\right\vert <\tilde{K}_{\pm }^{\,},
\label{e31.1}
\end{equation}%
and%
\begin{equation}
E_{\pm }^{\left( -\right) }\simeq \pm \left[ \tilde{K}_{\mp }^{\,}-\frac{1}{2%
}\zeta \tilde{K}_{\mp }^{\,2}+O\left( \zeta ^{2}\right) \right] \Rightarrow
\left\vert E_{\pm }^{\left( -\right) }\right\vert <\tilde{K}_{\mp }^{\,}.
\label{e32.1}
\end{equation}%
Moreover, the logarithmic nature of the solutions (\ref{e31}) and (\ref{e32}) (manifested by the exponential structure of the rainbow function here) as well as Figures 4(a) and 4(b), suggest that there are no eminent upper limits for the energies of both particles and anti-particles toward the Planck energy scale $E_{p}$.
\begin{figure}[!ht]  
\centering
\includegraphics[width=0.35\textwidth]{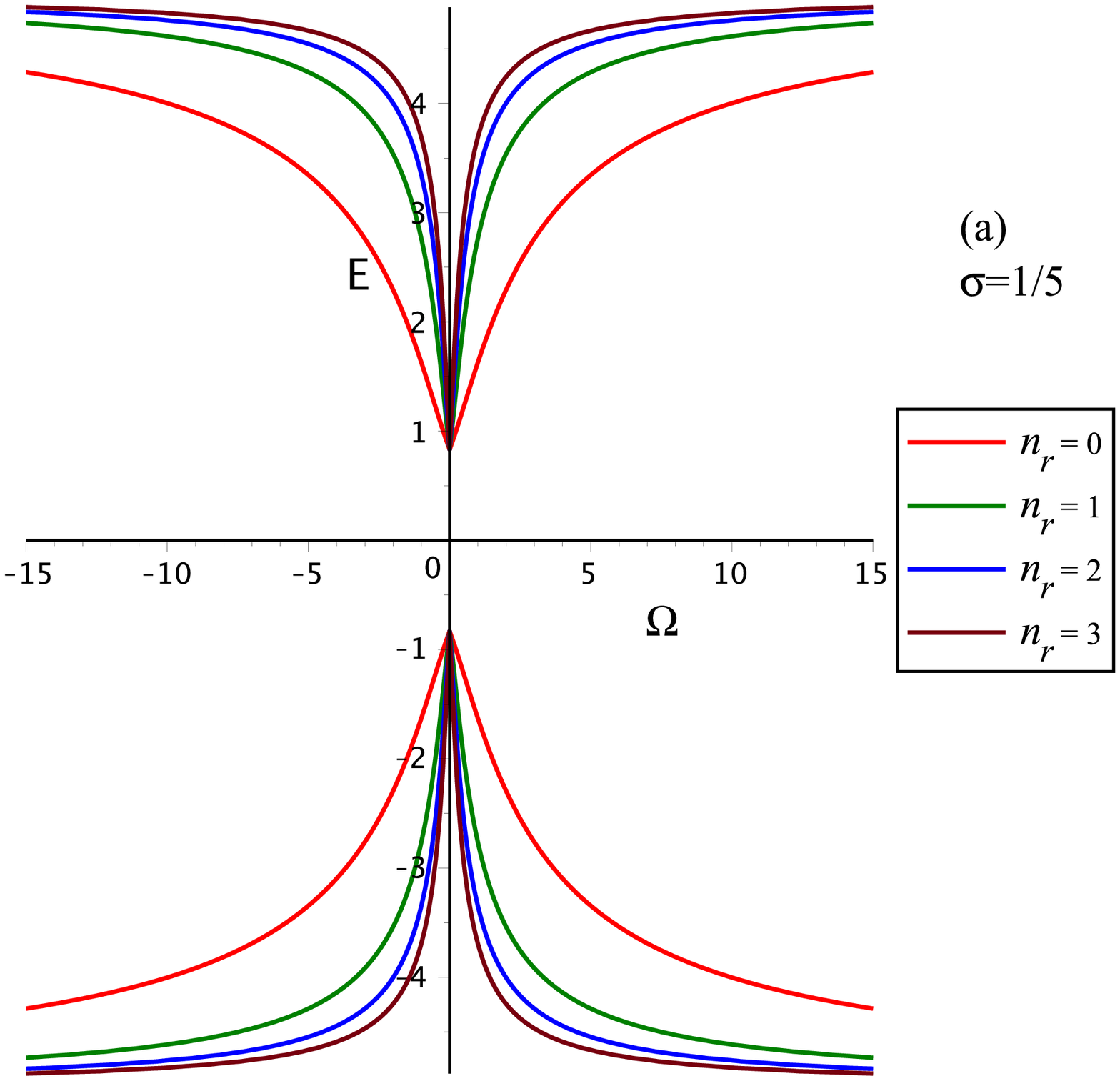}
\includegraphics[width=0.35\textwidth]{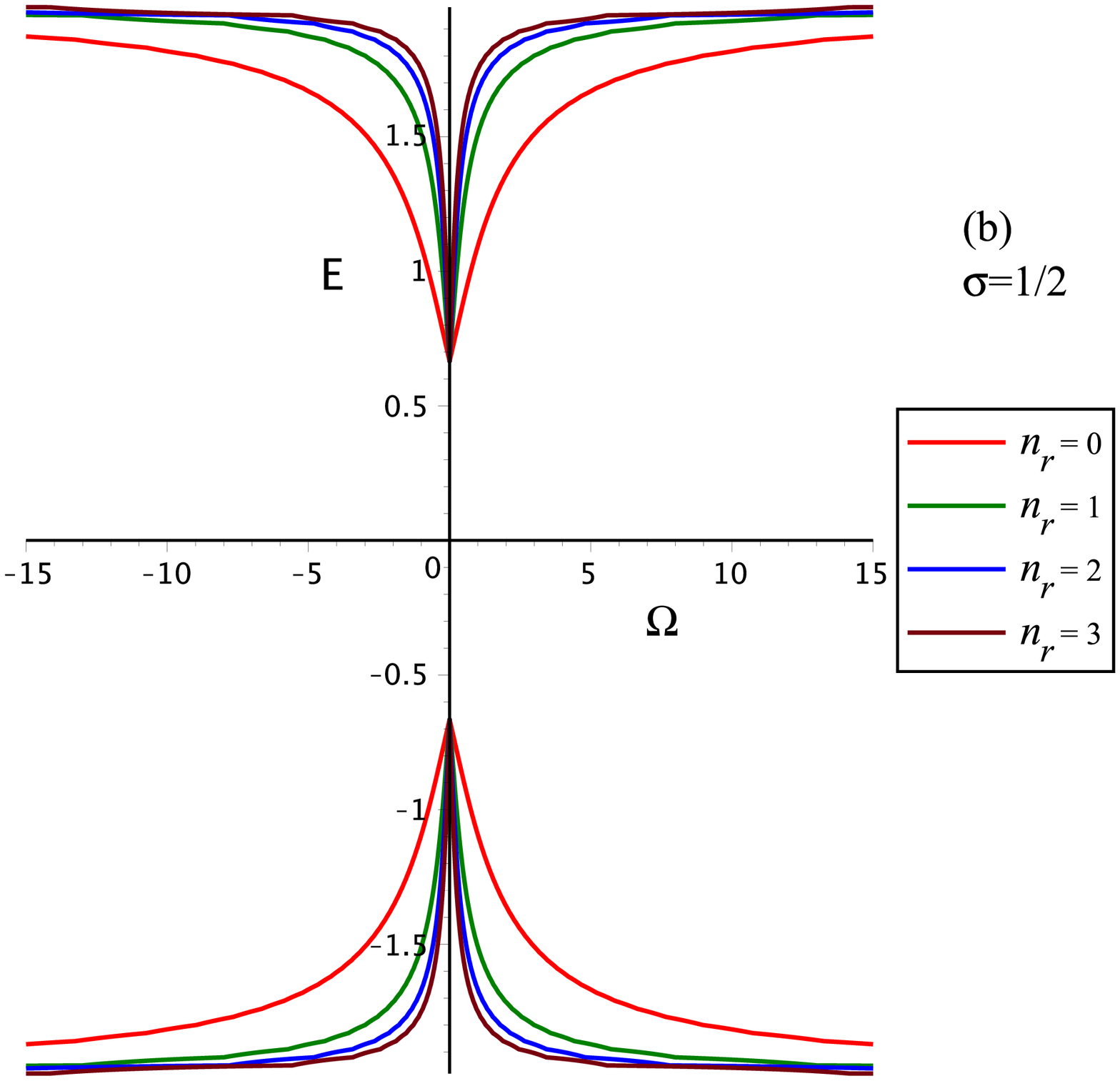} 
\caption{\small 
{ The energy levels of (\ref{e33}) and (\ref{e34}) for $%
g_{_{0}}\left( y\right) =\left( 1-\sigma \left\vert E\right\vert \right)
^{-1};\,\sigma =\epsilon /E_{p}$, $\,g_{_{1}}\left( y\right) =1$, using $%
\alpha =0.5$, $k_{z}=1$, $\ell =0$, and $n_{r}=0,1,2,3$ so that (a) shows $E$
against $\Omega $ for $\sigma =0.2$, and (b) shows $E$ against $\Omega $ for 
$\sigma =0.5$.}}
\label{fig5}
\end{figure}%

\subsection{Rainbow functions $g_{_{0}}\left( y\right) =\left( 1-%
\epsilon y\right) ^{-1},\,g_{_{1}}\left( y\right) =1$}

We have reported above that for the case $g_{_{0}}\left( y\right)
=g_{_{1}}\left( y\right) $, equations (\ref{e22}) and (\ref{e23}) imply no
rainbow gravity effect on the spectroscopic structure of massless KG-oscillators at hand. Therefore, the rainbow functions $g_{_{0}}\left(
y\right) =g_{_{1}}\left( y\right) =\left( 1-\epsilon y\right) ^{-1}$, used to resolve the horizon problem \cite{Re5,R13}, would have no effect on the spectra. However, it could be interesting to report that the rainbow
functions $g_{_{0}}\left( y\right) =\left( 1-\sigma \left\vert E\right\vert
\right) ^{-1};\,\sigma =\epsilon /E_{p}$, $\,g_{_{1}}\left( y\right) =1$ would yield interesting results that comply with the rainbow gravity model and preserve the invariance of the Planck's energy scale $E_{p}$. Such a
rainbow functions pair would yield that%
\begin{equation}
E_{\pm }^{\left( +\right) }=\pm \left( 1-\sigma \left\vert E\right\vert
\right) \tilde{K}_{\pm }^{\,}\Longrightarrow \left\{ 
\begin{tabular}{l}
$E_{+}^{\left( +\right) }=+\left( 1-\sigma E_{+}^{\left( +\right) }\right) 
\tilde{K}_{+}^{\,}$ \\ 
$E_{-}^{\left( +\right) }=-\left( 1+\sigma E_{-}^{\left( +\right) }\right) 
\tilde{K}_{-}^{\,}$%
\end{tabular}%
\right\} \Longrightarrow E_{\pm }^{\left( +\right) }=\pm \frac{\tilde{K}%
_{\pm }^{\,}}{1+\sigma \tilde{K}_{\pm }^{\,}},  \label{e33}
\end{equation}%
by (\ref{e22}) and 
\begin{equation}
E_{\pm }^{\left( -\right) }=\pm \left( 1-\sigma \left\vert E\right\vert
\right) \tilde{K}_{\mp }^{\,}\Longrightarrow \left\{ 
\begin{tabular}{l}
$E_{+}^{\left( -\right) }=+\left( 1-\sigma E_{+}^{\left( -\right) }\right) 
\tilde{K}_{-}^{\,}$ \\ 
$E_{-}^{\left( -\right) }=-\left( 1+\sigma E_{-}^{\left( -\right) }\right) 
\tilde{K}_{+}^{\,}$%
\end{tabular}%
\right\} \Longrightarrow E_{\pm }^{\left( -\right) }=\pm \frac{\tilde{K}%
_{\mp }^{\,}}{1+\sigma \tilde{K}_{\mp }^{\,}}  \label{e34}
\end{equation}%
by (\ref{e23}). One should notice that for $\left\vert \Omega \right\vert
\rightarrow \infty $ the energies $\left\vert E_{\pm }^{\left( \pm \right)
}\right\vert \sim 1/\sigma +O\left( 1/\Omega ^{2}\right) =E_{p}/\epsilon $.
Again this result shows that the rainbow parameter should satisfy $\epsilon
\geq 1$.

In Figures 5(a) and 5(b) the energy levels are shown for $\sigma =1/5$ and $%
\sigma =1/2$ for different values of the vorticity parameter $\Omega $. Figures 5(a) and 5(b) clearly show the for $\sigma =1/5$ the maximum energy
is $\left\vert E_{\pm }^{\left( \pm \right) }\right\vert \sim 5$ and for $%
\sigma =1/2$ the maximum energy is $\left\vert E_{\pm }^{\left( \pm \right)
}\right\vert \sim 2$. Moreover, the energy gap narrows as $\sigma $ increases. Yet an expansion about $\sigma =0$ would yield 
\begin{equation}
\left\vert E_{\pm }^{\left( +\right) }\right\vert =\tilde{K}_{\pm
}^{\,}-\sigma \tilde{K}_{\pm }^{\,2}+O\left( \sigma ^{2}\right) <\left\vert
E_{\pm }^{\left( +\right) }\right\vert _{\sigma =0}=\tilde{K}_{\pm }^{\,},
\label{e35}
\end{equation}%
and 
\begin{equation}
\;\left\vert E_{\pm }^{\left( -\right) }\right\vert =\tilde{K}_{\mp
}^{\,}-\sigma \tilde{K}_{\mp }^{\,2}+O\left( \sigma ^{2}\right) <\left\vert
E_{\pm }^{\left( -\right) }\right\vert _{\sigma =0}=\tilde{K}_{\mp }^{\,}.
\label{e36}
\end{equation}
Interestingly, this new experimental pair reproduces the same rainbow gravity effects as those used in loop quantum gravity \cite{R41,R42} (i.e., $g_{_{0}}\left(
y\right) =1$, $g_{_{1}}\left( y\right) =\sqrt{1-\epsilon E^{j}/E_{p}^{j}}%
;\,j=1,2$), That is, it preserves the invariance of the Planck's energy scale $E_{p}$ as well as symmetrization of the energy levels about $%
E=0$ for both KG-particles and anti-particles.

\section{Massless KG-particles in cosmic string spacetime and magnetic fields in a rainbow gravity introduced by $g_{_{0}}\left( y\right) =\left( 1-%
\epsilon y\right) ^{-1},\,g_{_{1}}\left( y\right) =1$}

In this section, we wish to show that the effects of rainbow gravity introduced by the new rainbow functions pair $g_{_{0}}\left( y\right)
=\left( 1-\epsilon y\right) ^{-1},\,g_{_{1}}\left( y\right) =1$, are as good as those introduced by the loop quantum gravity \cite{R41,R42} (i.e., $%
g_{_{0}}\left( y\right) =1$, $g_{_{1}}\left( y\right) =\sqrt{1-\epsilon
\left\vert E\right\vert ^{j}\text{/}E_{p}^{j}};\,j=1,2$), in the sense this pair preserves the invariance of the Planck's energy scale $E_{p}$ and the symmetry of the energy levels about $E=0$ for both KG-particles and anti-particles. Hereby, we consider massless KG-particles in cosmic string
rainbow gravity (i.e., $\Omega =0$, no vorticity) and in magnetic fields so that the KG-equation would read%
\begin{equation}
\frac{1}{\sqrt{-g}}\left( \partial _{\mu }-ieA_{\mu }\right) \sqrt{-g}g^{\mu
\nu }\left( \partial _{\upsilon }-ieA_{\upsilon }\right) \,\Psi =0,
\label{e37}
\end{equation}%
where we set $\Omega =0$ in Som-Raychaudhuri cosmic string rainbow gravity spacetime metric (\ref{eq2}) so that it reduces into just a cosmic string rainbow gravity spacetime metric. In a straightforward manner, one may follow the same steps as in section 2, with $A_{\mu }=\left( 0,0,A_{\varphi
},0\right) $, and obtain%
\begin{equation}
\left\{ \mathcal{E}^{2}+g_{_{1}}\left( y\right) ^{2}\left[ \partial _{r}^{2}+%
\frac{1}{r}\partial _{r}-\frac{\left( \ell -eA_{\varphi }\right) ^{2}}{%
\alpha ^{2}r^{2}}\right] \right\} \Psi \left( r\right) =0,  \label{e38}
\end{equation}%
with 
\begin{equation}
\mathcal{E}^{2}=g_{_{0}}\left( y\right) ^{2}E^{2}-g_{_{1}}\left( y\right)
^{2}k_{z}^{2}.  \label{e39}
\end{equation}%

At this point, one should notice that $A_{\varphi }=\frac{1}{2}B_{\circ
}r^{2}$ would result in a non-uniform magnetic field  $\mathbf{B=\nabla
\times A}=\frac{3}{2}B_{\circ }r\mathbf{\,}\hat{z}$, whereas $A_{\varphi }=%
\frac{1}{2}B_{\circ }r$ \ results in a uniform magnetic 
field $\mathbf{%
B=\nabla \times A}=\frac{1}{2}B_{\circ }\mathbf{\,}\hat{z}$. Whilst the former, in (\ref{e38}), would yield KG-oscillators%
\begin{equation}
\left\{ \partial _{r}^{2}-\frac{\left( \tilde{\ell}^{2}-1/4\right) }{r^{2}}+%
\frac{1}{4}\tilde{B}^{2}r^{2}+\Lambda _{Osc.}\right\} R\left( r\right)
=0;\;\Lambda _{Osc.}=\frac{\mathcal{E}^{2}+g_{_{1}}\left( y\right) ^{2}%
\tilde{\ell}\tilde{B}}{g_{_{1}}\left( y\right) ^{2}},\,\tilde{B}=\frac{%
eB_{\circ }}{\alpha }  \label{e40}
\end{equation}%
the later yields KG-Coulombic particles%
\begin{equation}
\left\{ \partial _{r}^{2}-\frac{\left( \tilde{\ell}^{2}-1/4\right) }{r^{2}}+%
\frac{\tilde{\ell}\tilde{B}}{r}+\Lambda _{Coul.}\right\} R\left( r\right)
=0;\;\Lambda _{Coul.}=\frac{\mathcal{E}^{2}-g_{_{1}}\left( y\right) ^{2}%
\tilde{B}^{2}/4}{g_{_{1}}\left( y\right) ^{2}}.  \label{e41}
\end{equation}%
Both are textbook Schr\"{o}dinger like problems and their exact solutions
are given by%
\begin{equation}
\Lambda _{Osc.}=\left\vert \tilde{B}\right\vert \left( 2n_{r}+\left\vert 
\tilde{\ell}\right\vert +1\right) ,\text{ \ and  }\Lambda _{Coul.}=-\frac{%
\tilde{\ell}^{2}\tilde{B}^{2}}{4\left( n_{r}+\left\vert \tilde{\ell}%
\right\vert +1/2\right) ^{2}}.  \label{e42}
\end{equation}

Under such settings, equation (\ref{e40}) would, for $g_{_{0}}\left(
y\right) =\left( 1-\sigma \left\vert E\right\vert \right) ^{-1},$ $\sigma
=\epsilon /E_{p},\,$\ $g_{_{1}}\left( y\right) =1$,  yield%
\begin{equation}
\frac{E_{osc.}^{2}}{\left( 1-\sigma \left\vert E\right\vert _{osc.}\right)
^{2}}=\mathcal{B}_{n_{r},\ell },\;\mathcal{B}_{n_{r},\ell }=\left\vert 
\tilde{B}\right\vert \left( 2n_{r}+\left\vert \tilde{\ell}\right\vert
+1\right) -\tilde{\ell}\tilde{B}+k_{z}^{2},  \label{e43}
\end{equation}%
\begin{figure}[!ht]  
\centering
\includegraphics[width=0.35\textwidth]{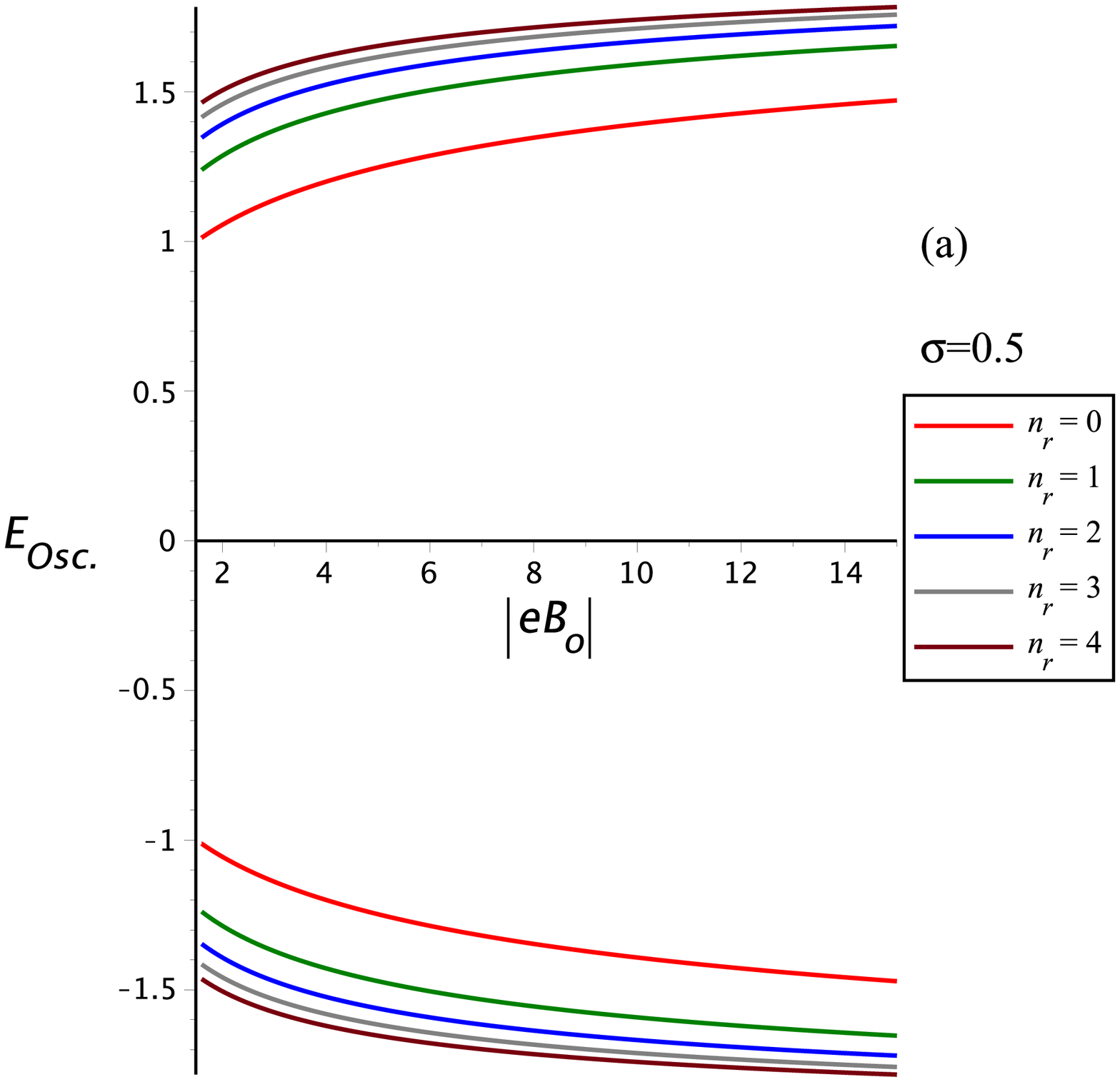}
\includegraphics[width=0.35\textwidth]{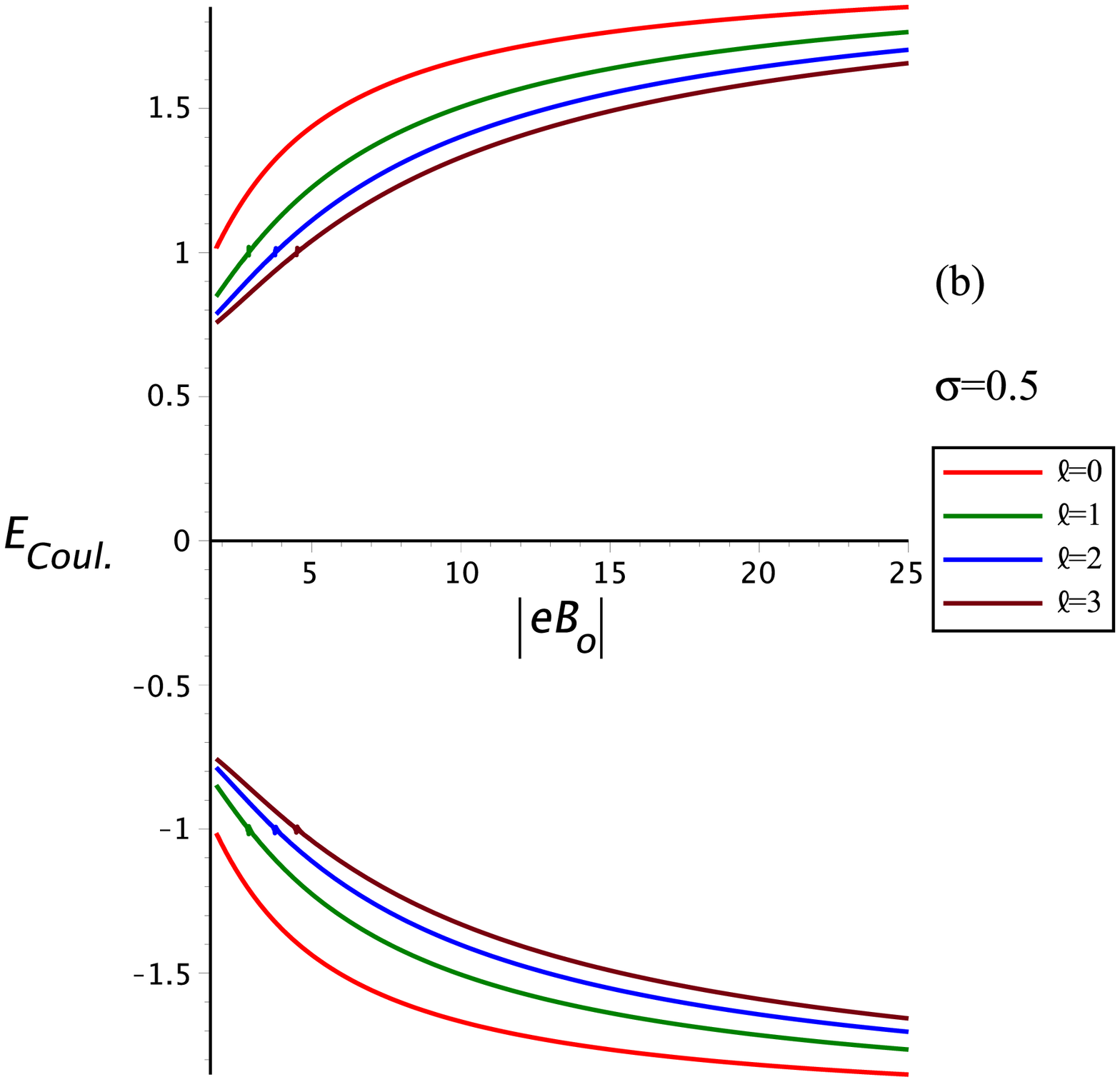} 
\caption{\small 
{ The energy levels for $g_{_{0}}\left( y\right) =\left(
1-\sigma \left\vert E\right\vert \right) ^{-1};\,\sigma =\epsilon /E_{p}$, $%
\,g_{_{1}}\left( y\right) =1$, using $\alpha =0.5$, $k_{z}=1$, $\sigma =0.5$
and at different values of  $\left\vert eB_{\circ }\right\vert $ so that (a)
shows the KG-oscillator energies (\ref{e43}) for $\ell =0$, $n_{r}=0,1,2,3,4$%
, and (b) shows the KG-Coulombic particles energies  (\ref{e44}) for $n_{r}=0$%
, $\ell =0,1,2,3.$.}}
\label{fig6}
\end{figure}%
and equation (\ref{e41}) gives%
\begin{equation}
\frac{E_{Coil.}^{2}}{\left( 1-\sigma \left\vert E\right\vert _{Coul.}\right)
^{2}}=\mathcal{C}_{n_{r},\ell },\;\mathcal{C}_{n_{r},\ell }=\frac{\tilde{B}%
^{2}}{4}\left( 1-\frac{\tilde{\ell}^{2}}{4\left( n_{r}+\left\vert \tilde{\ell%
}\right\vert +1/2\right) ^{2}}\right) +k_{z}^{2}.  \label{e44}
\end{equation}%
Obviously, the two results can be rewritten as%
\begin{equation}
E^{2}\left( 1-\sigma ^{2}\mathcal{K}\right) +2\sigma \mathcal{K}\left\vert
E\right\vert -\mathcal{K}=0\Leftrightarrow E_{\pm }^{2}\left( 1-\sigma ^{2}%
\mathcal{K}\right) \pm 2\sigma \mathcal{K}E_{\pm }-\mathcal{K}=0;\,E_{\pm
}=\pm \left\vert E\right\vert ,  \label{e45}
\end{equation}%
where $\mathcal{K}=\mathcal{B}_{n_{r},\ell }$ for KG-oscillators and $%
\mathcal{K}=\mathcal{C}_{n_{r},\ell }$ for KG-Coulombic particles. Such a
quadratic equation admits solution in the form of%
\begin{equation}
E_{\pm }=\frac{\mp \sigma \mathcal{K\pm }\sqrt{\mathcal{K}}}{1-\sigma ^{2}%
\mathcal{K}}.  \label{e46}
\end{equation}%
Consequently, an expansion about $\sigma =0$ would yield that $\left\vert
E_{\pm }\right\vert \approx \sqrt{\mathcal{K}}-\sigma \mathcal{K}%
^{2}+O\left( \sigma ^{2}\right) <\sqrt{\mathcal{K}}$ (where $\sqrt{\mathcal{K%
}}$ is the exact eigenvalue when rainbow gravity is switched off at $%
\epsilon =0$). Moreover, at the limit $\tilde{B}\rightarrow \infty $ we obtain  $\left\vert E_{\pm }\right\vert =1/\sigma =E_{p}/\epsilon
\Rightarrow \epsilon \geq 1$.

In Figures 6(a) and 6(b), we show the KG-oscillators' and KG-Coulombic particles' energies, respectively, at different $\left\vert eB_{\circ
}\right\vert $ values. Notably, at $\mathcal{K}=1/\sigma ^{2}$ of (\ref{e46}) we have energy states to fly away and disappear from the spectrum. We have, therefore, estimated and avoided such singularities while presenting our
results in Figures 6 (i.e., we have used $\left\vert eB_{\circ }\right\vert
\geq 1.6$ for 6(a) and $\left\vert eB_{\circ }\right\vert \geq 1.8$ for 6(b) ). In both cases, an obvious convergence to $\left\vert E_{\pm }\right\vert
=1/\sigma =2$ (for $\sigma =0.5$ used in the plots) is observed as $%
\left\vert eB_{\circ }\right\vert >>1.6$ in 6(a) and as $\left\vert
eB_{\circ }\right\vert >>1.8$ in 6(b). 

\section{Concluding remarks}

In the current study, we have considered massless KG-oscillators in Som-Raychaudhuri cosmic string rainbow gravity spacetime background. In the light of our observations above and within the fine tuning of the rainbow functions,  we give our concluding remarks as follows.

Among the rainbow functions we have used, we have observed the family of rainbow functions $g_{_{0}}\left( y\right) =1$, $%
g_{_{1}}\left( y\right) =\sqrt{1-\epsilon y^{j}}$; $j=1,2$ (used in loop quantum gravity) has fully delivered what rainbow gravity is designed for.  Such rainbow functions ensure that the energy levels (for massless KG-particles and anti-particles alike) are between $\pm mc^2$ and $\pm E_p$ so that $m c^2\leq |E|\leq E_p$. That is, the energies of the particles $E_+$ are within the limits   $m c^2\leq E_+\leq E_p$ and those for the anti-particles $E_-$ are within $-E_p\leq E_-\leq -mc^2$.  This would secure the invariance of the Planck's energy scale $E_p$ and consequently would justify our fine tuning of the rainbow functions (i.e., $y=E/E_p$ should be replaced by $y=|E|/E_p$), therefore. The rainbow functions pair $g_{_{0}}\left(
y\right) =g_{_{1}}\left( y\right) =\left( 1-\epsilon y\right) ^{-1}$ (used in the horizon problem) had no effects on the energy levels of the massless KG-oscillators as clearly suggested by equations (\ref{e22}) and (\ref{e23}). However, the pair $g_{_{0}}\left( y\right) =\left(
e^{\epsilon y}-1\right) /\epsilon y$, $g_{_{1}}\left( y\right) =1$ (a  gamma-ray bursts byproduct) has the effect of slowing down the migration of the energy levels towards infinity as may obviously be obtained by a simple comparison between Fig. 1(a) and 4(a) and 4(b). This is attributed to the exponential nature of this pair that manifestly yields a logarithmic nature of the energies as reported in (\ref{e31}) and (\ref{e32}).

In connection with the new rainbow functions pair $g_{_{0}}\left( y\right) =\left( 1-%
\epsilon y\right) ^{-1},\,g_{_{1}}\left( y\right) =1$, we have to emphasis that this is an experimental (metaphorically speaking) toy-model that has been accidentally discovered. This toy-model has shown consistent performance in terms of enforcing the energies of the particles $E_+$ to be contained within   $m c^2\leq E_+\leq E_p$ and that of the anti-particles $E_-$ within $-E_p\leq E_-\leq -mc^2$. Hence, it secures the invariance of the Planck's energy scale $E_p$. This is observed not only for the massless KG-oscillators in Som-Raychaudhuri cosmic string rainbow gravity spacetime (documented in section 2) but also for Massless KG-particles in cosmic string rainbow gravity spacetime and in a non-uniform and uniform magnetic fields (documented in section 3). Such encouraging and interesting performance should, in our opinion, stimulate thorough investigations on the validity and reliability of the proposed experimental rainbow functions toy-model, which readily lies far beyond the scope of the current study. To the best of our knowledge, the current study did not appear elsewhere.

\textbf{Data availability statement:} 
The authors declare that the data supporting the findings of this study are available within the paper. 

\textbf{Declaration of interest:}
The authors declare that they have no known competing financial interests or personal relationships that could have appeared to influence the work reported in this paper.

\end{document}